\documentclass[acmsmall,screen]{acmart}
\AtBeginDocument{%
  }
\usepackage{multirow}
\usepackage{arydshln}
\usepackage{algorithm}
\usepackage{algorithmic}
\usepackage{enumitem}
\usepackage{threeparttable}

\begin{document}

\title{UniFAR: A Unified Facet-Aware Retrieval Framework for Scientific Documents}

\author{Zheng Dou}
\email{miracle\_dz@buaa.edu.cn}
\affiliation{
  \institution{School of Computer Science and Engineering, Beihang University}  
  \city{Beijing}
  \country{China}
}
\author{Zhao Zhang}
\email{zhao\_zhang@buaa.edu.cn}
\affiliation{
  \institution{School of Computer Science and Engineering, Beihang University}  
  \city{Beijing}
  \country{China}
}
\author{Deqing Wang}
\email{dqwang@buaa.edu.cn}
\affiliation{
  \institution{School of Computer Science and Engineering, Beihang University}  
  \city{Beijing}
  \country{China}
}
\author{Yikun Ban}
\email{yikunb@buaa.edu.cn}
\affiliation{
  \institution{School of Computer Science and Engineering, Beihang University}  
  \city{Beijing}
  \country{China}
}

\author{Fuzhen Zhuang}
\email{zhuangfuzhen@buaa.edu.cn}
\affiliation{
  \institution{School of Artificial Intelligence, Beihang University}  
  \city{Beijing}
  \country{China}
}

\renewcommand{\shortauthors}{Dou et al.}

\begin{abstract}
  Scientific document retrieval (SDR) plays a critical role in modern scientific research, supporting knowledge discovery and evidence-based reasoning. It has evolved along two paradigms: document--document (doc-doc) retrieval driven by inter-document contrastive learning, and question--document (q-doc) retrieval emerging from LLMs and RAG for natural-language interaction. In practice, scientific workflows rely on both paradigms, requiring the retrieval of related papers given a seed document or identifying relevant documents given a user question. However, existing methods typically treat two paradigms separately, hindering their complementary strengths.
  To address this, we propose UniFAR, a unified facet-aware retrieval framework that jointly supports both doc-doc and q-doc retrieval within a shared representation space. 
  UniFAR introduces a multi-granularity representation and aggregation module to unify the encoding of {scientific papers and natural-language questions}, and a facet-level modeling mechanism with learnable anchors to {produce representations for three interpretable semantic facets---background, method, and result}.
  A facet-aware joint training strategy is further proposed to integrate doc-doc and q-doc contrastive objectives with sentence-level facet alignment, enabling unified learning from both inter-document relations and question-oriented supervision.
  {Experiments on five benchmarks spanning computer science, medicine, and biology demonstrate the effectiveness and generalizability of UniFAR in both doc-doc and q-doc retrieval tasks and across multiple backbone models.}
\end{abstract}




\maketitle
\section{Introduction}
\label{intro}
With the rapid growth of scientific publications worldwide, the ability to retrieve relevant scientific literature has become essential to knowledge discovery, evidence synthesis, and AI-assisted research \citep{ai4r,llm4research}. 
Recent scientific assistants and retrieval-augmented generation (RAG) \citep{RAG} systems already rely on retrieval over massive scholarly corpora, making \textbf{scientific document retrieval (SDR)} a central component of modern scientific workflows \citep{OpenScholar,OpenResearcher,SciRAG}. 
To support such large-scale information retrieval, dense retrieval \citep{DPR} {has become a widely adopted approach}, which shifts the focus from lexical to semantic matching by representing queries and documents in a shared embedding space. 
Early approaches rely on contextual encoders trained with language modeling objectives \citep{bert,mpnet}, while later developments adopt contrastive learning for improved representation quality \citep{sbert,declutr,SimCSE}.
More recently, advances have moved toward large-scale contrastive training with weak or self-supervision \citep{Contriever,E5}, leading to the emergence of retrievers that perform strongly across diverse retrieval tasks \citep{BEIR,MTEB}. 

{Scientific papers contain specialized terminology, and determining their relevance often requires modeling methodological and evidential relationships beyond overall topical similarity.
These characteristics motivate retrieval methods specifically tailored to scientific content.}
Early efforts focus on adapting pre-trained language models to scientific corpora, aiming to better capture domain-specific terminology and contextual semantics \citep{scibert,biobert}. To incorporate inter-document signals, subsequent approaches adopt citation-based contrastive learning, leveraging citation relations as supervision to learn representations that encode scholarly relatedness \citep{specter,SciNCL}. Building upon this, later work further exploits citation graph structures to incorporate richer contextual signals beyond pairwise citation links, such as methodological inspiration \citep{MIR} and empirical findings \citep{beyondcitations}. Beyond document-level representations, some studies also explore finer-grained signals from co-citation sentences \citep{aspire} or discourse-aware facets \citep{FLeW,FaBle} to enable more nuanced semantic modeling.
These studies substantially advance SDR by integrating domain-specific semantics, inter-document relations, citation context, and fine-grained discourse structure.
{Nevertheless, much of this retrieval-oriented work emphasizes inter-document relatedness, with representations primarily optimized for retrieving related papers given a seed paper (\textbf{doc-doc retrieval}), rather than for matching natural-language questions to papers.}

With the rapid development of large language models (LLMs) \citep{RAG-LLM}, scientific question answering \citep{qasper,qasa,paperqa}, and literature-grounded RAG systems \citep{OpenScholar,OpenResearcher,SciRAG}, users increasingly access scientific corpora through natural-language questions instead of seed documents alone. As a result, SDR is further required to support question-driven retrieval, where the goal is to retrieve documents that answer a user question, provide evidence for a claim, or support multi-document synthesis  (\textbf{q-doc retrieval}). {Consequently, scientific retrieval is no longer only about modeling inter-document relatedness, but should also align with user intent and fine-grained needs.} This shift toward question-driven retrieval highlights the need for a unified scientific retrieval framework that supports both doc-doc and q-doc retrieval paradigms. 
These two paradigms are inherently complementary {in scientific workflows}: doc-doc retrieval captures scholarly relatedness, citation structure, and global research context, while q-doc retrieval focuses on user intent, fine-grained evidence, and task-specific information needs. 

\begin{figure*}[!t]
  \centering
  \includegraphics[width=0.99\linewidth]{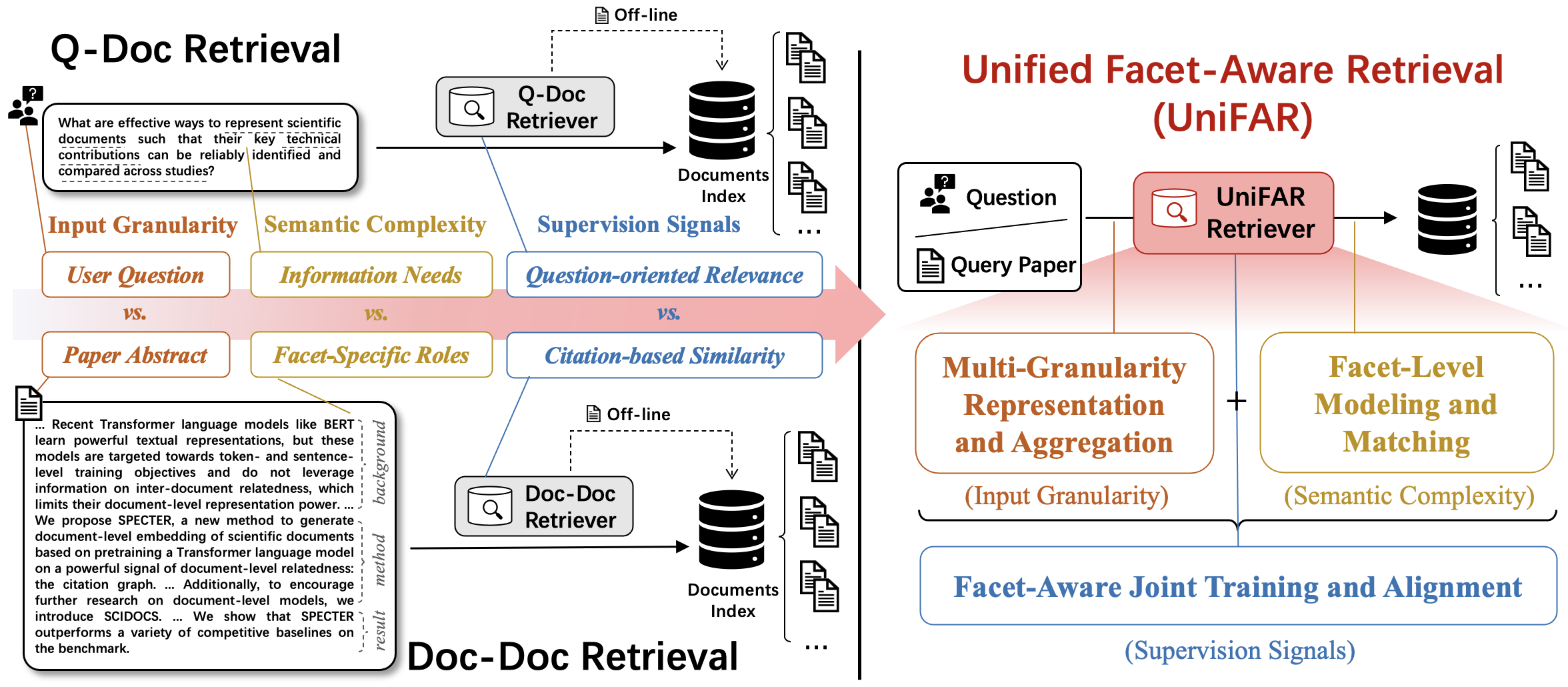}
  \caption{{Overview of the misalignments between q-doc and doc-doc retrieval (left) and the key components of UniFAR designed to address them (right). In our setting, each scientific paper is represented by its title and abstract. The labels in the example abstract illustrate the background, method, and result facets.}}
  \label{fig_intro}
\end{figure*}

{However, this complementarity is not explicitly exploited when the two paradigms are learned independently.} {A unified framework can reuse candidate paper representations within a shared space across paper and question queries, while joint modeling of inter-document relations and question-specific relevance can provide complementary signals for richer representations.}
{In established scientific retrieval methods \citep{specter,SciNCL}, each paper's title and abstract are used as its textual input, which provides a compact description of the research content.} {We follow this setting and characterize the systematic misalignments between the two retrieval paradigms along three dimensions, as illustrated in Figure~\ref{fig_intro} (left).
\emph{(1) Input Granularity}. Paper abstracts summarize different aspects of a study across multiple sentences, whereas questions may express focused information needs through a few key terms or combine several requirements across sentences. The information relevant to retrieval can therefore occur at different granularities, requiring appropriate aggregation of token-level details and sentence-level context across input types.
\emph{(2) Semantic Complexity}.
Scientific abstracts contain information with distinct semantic roles, commonly modeled through three facets \citep{FLeW,FaBle,csfcube}: background, describing the research context and problem; method, describing the approach; and result, describing the findings and outcomes.
These facets provide interpretable perspectives for comparing scientific content according to research problems, approaches, and findings.
User questions may target one or several of these facets, while an abstract also contains information outside the user's focus.
Consequently, overall topical similarity does not necessarily imply relevance to the requested facets, motivating explicit facet-level representation and matching.
\emph{(3) Supervision Signals}. Citation-based or document-similarity supervision captures relationships between scientific papers, but does not directly specify which papers are relevant to a particular user question.
Supporting both retrieval paradigms therefore requires integrating inter-document relationships with explicit question-oriented supervision.}

In this paper, we propose \textbf{UniFAR}, a \textbf{Uni}fied \textbf{F}acet-\textbf{A}ware \textbf{R}etrieval framework for SDR that jointly supports doc-doc and q-doc retrieval within a shared representation space. As illustrated in Figure~\ref{fig_intro} (right), UniFAR is designed to explicitly bridge the input, semantic, and supervision misalignments identified above through three key components.
{First, UniFAR introduces a \emph{multi-granularity representation and aggregation} module, which encodes scientific papers and natural-language questions in a unified format and adaptively aggregates token-level and sentence-level information. Second, it performs \emph{facet-level modeling and matching}, where learnable facet anchors associated with background, method, and result guide the construction of input-specific facet embeddings for facet-aware matching between queries and candidates.} Third, it adopts a \emph{facet-aware joint training and alignment} strategy, which jointly optimizes doc-doc and q-doc retrieval objectives while aligning facet-level attention with sentence-level facet supervision.
{The same three-facet scheme is used to organize the Facet-Aware Training Units (FTUs), define the facet anchors, and supervise facet--sentence alignment, connecting training data construction with representation learning.}
By addressing these misalignments in a unified framework, UniFAR enables effective scientific retrieval across both doc-doc and q-doc settings. The main contributions of this work are summarized as follows:
\begin{itemize}[label=---]
\item We formulate unified scientific retrieval as learning a shared representation space that supports both doc-doc and q-doc retrieval, and propose \textbf{UniFAR}, a unified facet-aware retrieval framework with multi-granularity representation, facet-level aggregation, and facet-aware matching.
\item We introduce a facet-aware joint training strategy for unified scientific retrieval. Specifically, we construct large-scale \textbf{Facet-Aware Training Units (FTUs)} and jointly optimize doc-doc and q-doc contrastive losses together with a facet-sentence alignment loss.
\item {We evaluate UniFAR across multiple backbone encoders on five
benchmarks from different fields, covering doc-doc and q-doc retrieval.} {In the instruction-based
evaluation, UniFAR with Qwen3Emb-8B achieves the highest nDCG@10
in every reported setting, outperforming both the single-vector and
facet-instructed variants of Qwen3Emb-8B and GritLM-7B.}
\end{itemize}

\section{Related Work}
\subsection{Document-Centric Representation Learning}
Representation learning maps queries and documents into a shared semantic space for similarity matching \citep{DPR}. 
In scientific document retrieval, existing work predominantly adopts a document-centric paradigm, where representations are optimized to capture inter-document relationships and scholarly context. 
These approaches can be broadly divided into document-level and fine-grained representation learning.

\textit{Document-Level.} Early work adapts pretrained language models such as SciBERT \citep{scibert} to scientific corpora, obtaining document embeddings via standard pooling. 
Building on this foundation, SPECTER \citep{specter} introduces citation-informed contrastive learning, where cited documents are treated as positives to encode scholarly relatedness. 
SciNCL \citep{SciNCL} further extends this paradigm by incorporating citation graph structure through neighborhood contrastive learning, enabling richer and more informative supervision signals. 
Following this paradigm, SPECTER-2 \citep{specter2} expands the training corpus and incorporates multiple task formats to improve generalization across downstream retrieval tasks, while SciMult \citep{scimult} introduces multi-task contrastive pretraining to learn more generalizable scientific representations across diverse training signals.
Despite their effectiveness, these methods encode each document into a single dense vector, which may limit their ability to capture fine-grained semantics within scientific texts.

\textit{Fine-Grained.} To address the limitations of whole-document representations, studies have shifted toward modeling fine-grained semantic structures within documents. 
MixGR \citep{mixgr} integrates multi-granularity signals by combining representations at different levels, enabling more flexible matching between queries and long documents. 
FaBle \citep{FaBle} explicitly models semantic facets and constructs facet-aware representations to support fine-grained relevance estimation. 
Beyond improving representation granularity, a line of work further transitions from single-vector to multi-vector representations. 
ASPIRE \citep{aspire} represents documents as sets of sentence embeddings and performs fine-grained matching through sentence-level interactions, while FLeW \citep{FLeW} constructs facet-specific embeddings based on citation intent, enabling structured modeling of scientific discourse. 
A recent theoretical work \citep{limitations} has identified inherent limitations of single-vector embedding-based retrieval, thus providing support for multi-vector approaches.
Overall, these methods improve fine-grained semantic modeling, but remain largely confined to document-centric supervision and retrieval settings.

\subsection{Question-Driven Retrieval in the Era of LLMs}
The development of large language models (LLMs) and retrieval-augmented generation (RAG) has introduced new retrieval settings centered on natural-language questions \citep{RAG,RAG-LLM}. In the scientific literature domain, question-driven retrieval tightly integrates with scientific reasoning and generation, serving as a key interface between users, language models, and large-scale scientific corpora.
For instance, OpenScholar \citep{OpenScholar} supports scientific literature synthesis by retrieving and aggregating evidence from multiple papers to generate citation-grounded content, while OpenResearcher \citep{OpenResearcher} integrates retrieval into end-to-end AI-assisted research workflows with query understanding, retrieval, and refinement modules. 
SciRAG \citep{SciRAG} further advances this direction by incorporating adaptive retrieval and citation-aware generation for structured scientific reasoning.

Alongside this shift, a growing body of work has focused on aligning retrieval with natural-language queries. 
Scientific question answering benchmarks such as QASPER \citep{qasper} and QASA \citep{qasa} formalize information-seeking queries grounded in research papers, while systems like PaperQA \citep{paperqa} demonstrate how retrieval can be tightly coupled with question answering to support evidence-based responses. 
To better capture diverse user information needs, recent approaches further explore query-centric supervision and expansion strategies, where methods such as CCQGen \citep{CCQGen} and CCExpand \citep{CCExpand} construct or augment query sets to improve semantic coverage and retrieval robustness. 
Despite these advances, question-driven retrieval is typically modeled independently from document-centric approaches, limiting the ability to leverage inter-document relationships.

\subsection{Towards Unified Retrieval}
The above two paradigms highlight complementary strengths in scientific document retrieval. 
Document-centric modeling excels at capturing inter-document relationships and scholarly discourse, while question-driven retrieval focuses on specific user intent and downstream information needs. 
However, these paradigms are largely developed in isolation, limiting their ability to be jointly exploited in practical retrieval scenarios.
Therefore, a unified retrieval framework is needed to bridge these paradigms and enable generalizable retrieval capabilities.

\section{Problem Formulation}
\label{sec:problem}
Let $\mathcal{D}$ denote a corpus of scientific documents and $\mathcal{Q}$ the query space.
{A standard single-vector dense retrieval model learns a representation function
$f_{\mathrm{dense}}: \mathcal{Q} \cup \mathcal{D} \rightarrow \mathbb{R}^h$
that maps queries and documents into a shared embedding space.}
Given a query $q \in \mathcal{Q}$ and a candidate document $c \in \mathcal{D}$,
their semantic relevance is measured by a similarity function:
\begin{equation}
s(q,c) = \mathrm{sim}\big(f_{\mathrm{dense}}(q), f_{\mathrm{dense}}(c)\big),
\end{equation}
where $\mathrm{sim}(\cdot)$ is typically instantiated as dot product or cosine similarity.
{The retrieval task is to identify the top-$k$ candidate documents according to the relevance score:}
\begin{equation}
\mathrm{TopK}(q) = \operatorname*{arg\,top\text{-}k}_{c \in \mathcal{D}} \ s(q,c).
\end{equation}

{In this work, we extend the standard dense retrieval formulation as follows.}
{First, we consider a unified query space where a query $q \in \mathcal{Q}$ can be either a scientific paper (doc-doc retrieval) or a natural-language question (q-doc retrieval).} {In our setting, a document refers to a scientific paper and constitutes the retrieval unit, while its title and abstract serve as the textual input for computing its representation.
This input convention applies to both query and candidate papers, while natural-language questions are encoded from their own text.}

{Second, we adopt three common semantic facets of scientific abstract content: \textit{background}, describing the research context and problem,} {including prior work when it contextualizes or motivates the study};
{\textit{method}, describing the approach, techniques, or procedures;
and \textit{result}, describing the findings and outcomes.}
{We denote this fixed facet set as $\mathcal{F} = \{\textit{bg}, \textit{mt}, \textit{rs}\}$.} {These facets describe semantic roles within abstracts rather than the section structure of full papers.
For natural-language questions, the same facets provide semantic perspectives on the information sought, which may concern one or several of these facets.}

Accordingly, instead of representing each input with a single embedding vector, we map each input $x \in \mathcal{Q} \cup \mathcal{D}$ to a set of facet-level embeddings:
\begin{equation}
f_{\mathrm{facet}}(x)
=
\{\mathbf{e}_x^{(f)}\}_{f \in \mathcal{F}},
\quad
\mathbf{e}_x^{(f)} \in \mathbb{R}^h.
\end{equation}
{The relevance score between a query and a candidate document is then computed from their facet-level representations:}
\begin{equation}
s(q,c) =
\phi\big(
\{\mathbf{e}_q^{(f)}\}_{f \in \mathcal{F}},
\{\mathbf{e}_c^{(f)}\}_{f \in \mathcal{F}}
\big),
\end{equation}
where $\phi(\cdot)$ denotes a facet-aware similarity function.
Our goal is to learn a unified and facet-aware retrieval model that represents queries and documents in a shared embedding space, enabling fine-grained alignment across query types and semantic facets.

\section{UniFAR Framework}
\label{sec:framework}

{In this section, we present UniFAR, a unified facet-aware retrieval framework that instantiates the formulation in Section~\ref{sec:problem}. 
UniFAR implements $f_{\mathrm{facet}}(\cdot)$ to produce representations for the three semantic facets---background, method, and result---and computes relevance through facet-aware matching.
As illustrated in Figure~\ref{fig_1}, the framework consists of two main components:
(1) a replaceable backbone model that produces contextual token-level and sentence-level representations, and
(2) a facet-level aggregation module that uses learnable facet anchors, multi-granularity attention, and weighted fusion to construct input-specific facet embeddings.}

\subsection{Input Encoding and Representation}
\label{encode}
To achieve unified modeling for both scientific documents and scientific questions, UniFAR adopts a consistent input structure and a replaceable backbone model for multi-granularity representation.
\subsubsection{{Input Format.}}
{Following the input convention in Section~\ref{sec:problem}, each scientific paper is encoded from its title and abstract, while a natural-language question is encoded from its own text.
We segment each abstract and question into sentences and organize the input as an ordered sequence of sentence units.
For paper inputs, the title is treated as an additional sentence unit preceding the abstract sentences.
Using BERT-style special tokens for illustration, the two input forms are:}
\begin{equation}
\begin{aligned}
\textit{Input}_{d}
&=
\textit{Title}~\texttt{[SEP]}~
S_1~\texttt{[SEP]}~\cdots~
\texttt{[SEP]}~S_{n_d},\\
\textit{Input}_{q}
&=
\texttt{[CLS]}~
S_1~\texttt{[SEP]}~\cdots~
\texttt{[SEP]}~S_{n_q},
\end{aligned}
\label{eq:input}
\end{equation}
{where $n_d$ is the number of abstract sentences and $n_q$ is the number of question sentences.}
{These forms illustrate the sentence organization, while the actual special tokens and tokenization follow the selected backbone.}
{This organization preserves sentence boundaries across both input types and supports subsequent token- and sentence-level aggregation.}

\subsubsection{{Multi-Granularity Representation.}}
\begin{figure*}[!t]
  \centering
  \includegraphics[width=0.6\linewidth]{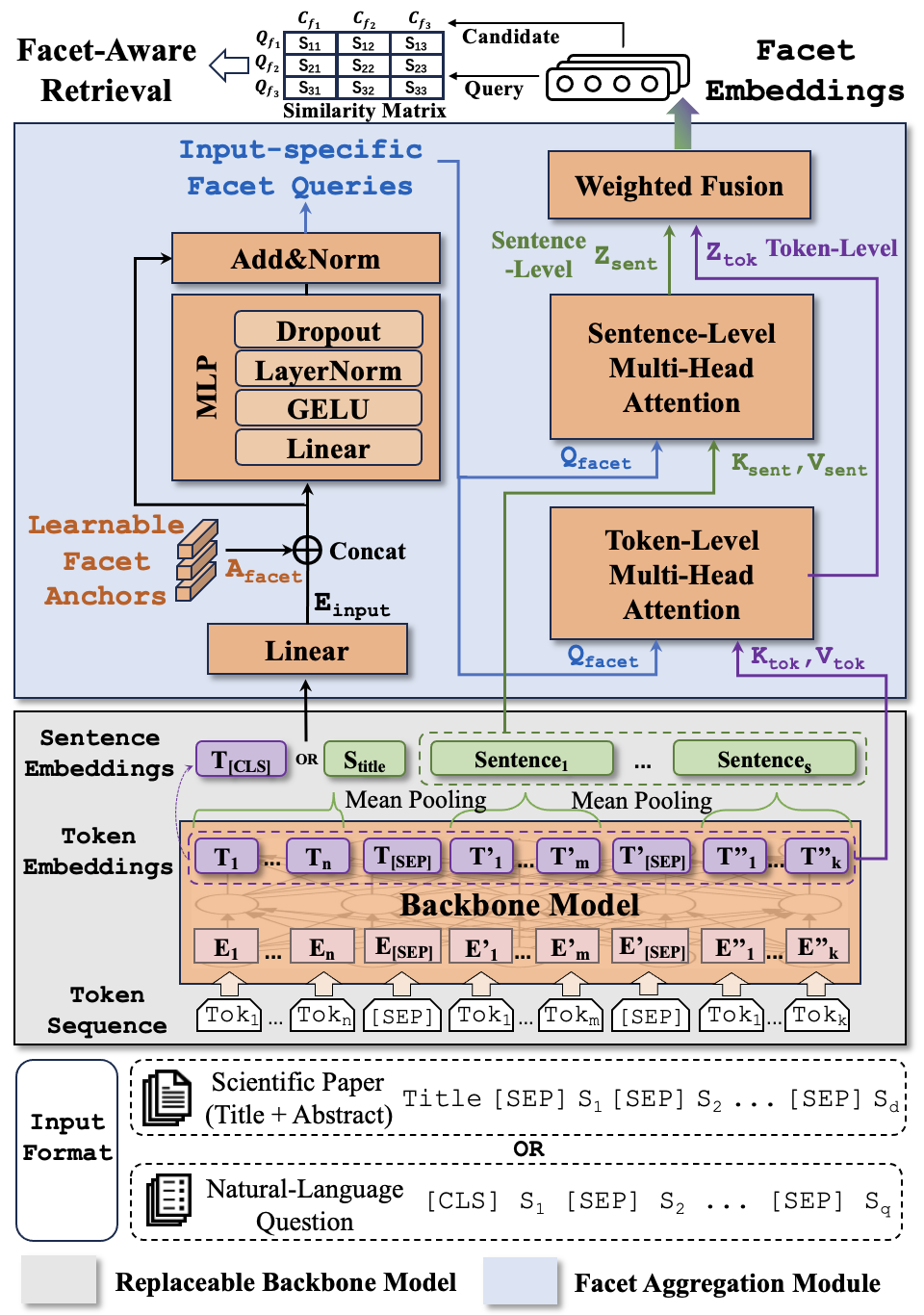}
  \caption{The architecture of our UniFAR framework. {Scientific papers} and questions are formatted as a unified sequence of segmented sentences and encoded by the \textbf{Replaceable Backbone Model (gray)} into token-level and sentence-level embeddings. \textbf{Facet Aggregation Module (blue)} builds on learnable facet anchors to produce input-specific facet queries, which guide the aggregation through multi-granularity attention. {Sentence-level and token-level attention produce corresponding facet representations, which are combined through weighted fusion to generate final facet-level embeddings.} Facet embeddings are used to compute a similarity matrix between query and candidate for each facet, supporting facet-aware retrieval.}
  \label{fig_1}
\end{figure*}

{As shown in Figure~\ref{fig_1} (gray), the replaceable backbone encodes an input sequence $x$ containing $L_x$ tokens into contextual token representations:}
\begin{equation}
\mathbf{T}
=
\operatorname{Backbone}(x)
\in\mathbb{R}^{L_x\times h},
\label{eq:token}
\end{equation}
{where $h$ is the backbone's hidden size and the $i$-th row, $\mathbf{T}_i\in\mathbb{R}^{h}$, represents the corresponding token.
Let $\mathcal{S}$ denote the sentence units in input $x$, with $n_x=|\mathcal{S}|$.
For a paper, these units comprise the title and abstract sentences; for a question, they comprise the sentences in the question text.
We obtain a representation for each sentence unit by mean pooling its contextual token embeddings:}
\begin{equation}
\mathbf{S}_i
=
\frac{1}{|\mathcal{I}_i|}
\sum_{j\in\mathcal{I}_i}\mathbf{T}_j,
\label{eq:sentence}
\end{equation}
{where $\mathcal{I}_i$ contains the token positions belonging to the $i$-th sentence unit.
The resulting sentence representations form
$\mathbf{S}\in\mathbb{R}^{n_x\times h}$.}

{For paper inputs, the title representation
$\mathbf{S}_{\textit{title}}\in\mathbb{R}^{h}$
is included in $\mathbf{S}$.
Thus, $n_x=n_d+1$ for a paper and $n_x=n_q$ for a question.
The token and sentence representation matrices provide the two granularities used by the subsequent facet aggregation module.}

\subsection{Facet-Level Aggregation}
\label{aggregation}
This module introduces learnable facet anchors to construct input-specific facet queries, which guide multi-granularity attention over sentence-level and token-level representations.
{A weighted fusion mechanism then combines the two granularity-specific representations of each facet to produce the final facet embeddings.}

\subsubsection{Facet Anchors and Input-Specific Queries.}
{We associate one learnable anchor
$\mathbf{a}^{(f)}\in\mathbb{R}^{h}$
with each facet
$f\in\mathcal{F}=\{\textit{bg},\textit{mt},\textit{rs}\}$.
The three anchors form
$\mathbf{A}_{\mathrm{facet}}\in\mathbb{R}^{3\times h}$
and serve as shared semantic reference points for representing papers and questions.}

{As illustrated in Figure~\ref{fig_1} (blue, left), each anchor is conditioned on input-specific context.
For paper inputs, we project the title representation
$\mathbf{S}_{\textit{title}}$
into a contextual vector
$\mathbf{E}_{\mathrm{input}}\in\mathbb{R}^{h}$.
For question inputs, we apply the same projection to a sequence-level summary obtained using the selected backbone's pooling convention.
For a backbone using \texttt{[CLS]} pooling, this summary is
$\mathbf{T}_{[\mathrm{CLS}]}$.
Each contextual vector is concatenated with the three facet anchors separately.
A lightweight MLP followed by residual addition and layer normalization transforms each concatenation into an input-specific facet query:}
\begin{equation}
\mathbf{q}_x^{(f)}
=
\operatorname{Add\&Norm}\!\left(
\operatorname{MLP}\!\left(
[\mathbf{a}^{(f)};\mathbf{E}_{\mathrm{input}}]
\right)
\right), \; f\in\mathcal{F}.
\label{eq:facet_query}
\end{equation}
{Here, $[\cdot;\cdot]$ denotes feature concatenation, and each resulting facet query
$\mathbf{q}_x^{(f)}\in\mathbb{R}^{h}$.
The three queries form an input-specific matrix
$\mathbf{Q}_{\mathrm{facet}}\in\mathbb{R}^{3\times h}$.}

{Conditioning the anchors on the input allows the facet queries to retain their shared semantic roles while adapting to the content being encoded.
These queries guide attention toward input information relevant to each facet.}

\subsubsection{Multi-Granularity Attention {and Fusion.}}
{Given the input-specific facet queries
$\mathbf{Q}_{\mathrm{facet}}$,
UniFAR defines two attention branches, as shown in Figure~\ref{fig_1} (blue, right).
The sentence and token representation matrices $\mathbf{S}\in\mathbb{R}^{n_x\times h}$ and $\mathbf{T}\in\mathbb{R}^{L_x\times h}$ serve as the key--value inputs to their respective multi-head attention modules:}
\begin{equation}
\begin{aligned}
\mathbf{Z}_{\mathrm{sent}}
&=
\operatorname{MHA}_{\mathrm{sent}}
\big(
\mathbf{Q}_{\mathrm{facet}},
\mathbf{S},
\mathbf{S}
\big),\\
\mathbf{Z}_{\mathrm{tok}}
&=
\operatorname{MHA}_{\mathrm{tok}}
\big(
\mathbf{Q}_{\mathrm{facet}},
\mathbf{T},
\mathbf{T}
\big),
\end{aligned}
\label{eq:attention}
\end{equation}
{where both
$\mathbf{Z}_{\mathrm{sent}}$ and $\mathbf{Z}_{\mathrm{tok}} \in \mathbb{R}^{3\times h}$, with one row per facet $f \in \mathcal{F}$.}

{The two branches capture facet-relevant information at different granularities.
Sentence-level attention aggregates contextualized sentence semantics, while token-level attention provides access to finer-grained evidence within the input.
}{We combine the two branches using three facet-specific fusion weights
$\mathbf{w}=(w_{\mathrm{bg}},w_{\mathrm{mt}},w_{\mathrm{rs}})$.
For each facet $f$, the final facet embedding is computed as:}
\begin{equation}
\mathbf{e}_x^{(f)}
=
w_f \mathbf{z}_{\mathrm{sent}}^{(f)}
+
(1-w_f)\mathbf{z}_{\mathrm{tok}}^{(f)},
\quad f\in\mathcal{F}=\{\textit{bg},\textit{mt},\textit{rs}\},
\label{eq:fusion}
\end{equation}
{where $\mathbf{z}_{\mathrm{sent}}^{(f)}$ and
$\mathbf{z}_{\mathrm{tok}}^{(f)}$ denote the rows associated with facet $f$
in $\mathbf{Z}_{\mathrm{sent}}$ and $\mathbf{Z}_{\mathrm{tok}}$, respectively. Each weight $w_f \in [0,1]$ controls the relative contribution of
the sentence-level and token-level branches for its corresponding
facet, yielding three facet embeddings $f_{\mathrm{facet}}(x)=\{\mathbf{e}_x^{(f)}\}_{f\in\mathcal{F}}$. Different fusion strategies are compared in Section~\ref{exp:architecture} to determine the configuration.}

\subsection{Facet-Aware Retrieval}
\label{retrieval}
UniFAR supports both retrieval paradigms by treating the query as either a {scientific paper} or a natural-language question, while each candidate is a {scientific paper} in the corpus.
{For a query $q$ and a candidate $c$, the model produces facet embeddings $\{\mathbf{e}_q^{(f)}\}_{f\in\mathcal{F}}$ and $\{\mathbf{e}_c^{(f)}\}_{f\in\mathcal{F}}$, respectively.}
{Candidate paper representations can be computed offline and reused across both query types.}

\subsubsection{Facet-Level Similarity Matrix.}
{We define a similarity matrix $\mathbf{M}_{q,c}\in\mathbb{R}^{3\times3}$, whose rows and columns correspond to the $background, method,$ and $result$ facets of the query and candidate, respectively.} Each entry measures the similarity between a query facet and a candidate facet:
\begin{equation}
\mathbf{M}_{q,c}^{(f,g)}
=
\operatorname{sim}\big(
\mathbf{e}_q^{(f)},\mathbf{e}_c^{(g)}
\big),
\quad f,g\in\{bg,mt,rs\}.
\label{eq:facet_similarity}
\end{equation}
{The diagonal entries compare corresponding semantic facets, while the off-diagonal entries compare different facets.
This matrix provides a common basis for specifying and comparing facet-matching strategies.}

\subsubsection{Facet-Aware Relevance Scoring.}
\label{sec:facet_scoring}
We compute the overall relevance score by aggregating the entries of the facet similarity matrix:
\begin{equation}
s(q,c)=\phi(\mathbf{M}_{q,c}),
\label{eq:facet_scoring}
\end{equation}
{where $\phi(\cdot)$ specifies how facet-level similarities contribute to retrieval relevance. This design generalizes standard dense retrieval, where relevance is computed from a single embedding pair, to a structured matching in the matrix over multiple semantic facets.
As a general-purpose default, we average the diagonal entries,
$\phi(\mathbf{M}_{q,c})=\frac{1}{3}\sum_{f\in\mathcal{F}}
\mathbf{M}_{q,c}^{(f,f)}$.
This strategy preserves alignment between corresponding facets
and assesses overall similarity across the three semantic roles.}
{Alternative facet-matching strategies are evaluated and analyzed in Section~\ref{sec:matching}.}

\section{Facet-Aware Joint Training}
\label{sec:training}

In this section, we describe the facet-aware joint training strategy of UniFAR. 
It complements the \textit{static} representation and computation mechanisms in Section~\ref{sec:framework} by defining how data and gradients flow \textit{dynamically} through the model to optimize parameters. 
As illustrated in Figure~\ref{fig_2}, the training pipeline consists of three stages: (1) training data construction with Facet-Aware Training Units (FTUs), (2) forward encoding with UniFAR, and (3) joint optimization with multiple objectives.

\subsection{Training Data} 
{To support joint learning of doc-doc and q-doc retrieval, we construct training data in the form of Facet-Aware Training Units (FTUs), which associate facet-specific paper relations with sentence-level facet labels and generated scientific questions.}

\subsubsection{{Data Construction.}}
\label{data construction}
{We construct FTUs from the facet-specific triplets in FLeW \citep{FLeW}, where each triplet $(d_q, d_{+}^{(f)},d_{-}^{(f)})$ associates a query paper with a positive and a negative paper for facet $f\in\{\textit{bg},\textit{mt},\textit{rs}\}$.}

{First, we screen all source papers against
the query and candidate papers in the evaluation benchmarks (Section~\ref{benchmarks}), using paper identifiers, normalized DOIs, and normalized titles.
If a query paper overlaps with a benchmark paper, all its associated triplets are removed; if a positive or negative paper overlaps, only triplets containing that paper are removed.}
{We deduplicate the remaining triplets and group them by query paper $d_q$, forming facet-specific positive and negative sets, $\mathcal{D}_{\mathrm{pos}}^{(f)}$ and $\mathcal{D}_{\mathrm{neg}}^{(f)}$ for each facet $f$.} {Only groups with non-empty positive and negative sets for all three facets are retained.}

{Then, we conduct LLM-based annotation and generation to produce sentence-level facet labels and facet-aware questions} {using \texttt{Qwen3-Max-2025-09-23}}. {Following our settings, all papers are represented by their titles and abstracts.} {We prompt the LLM to assign each abstract sentence to one of the three facets $f\in\{\textit{bg},\textit{mt},\textit{rs}\}$, obtaining sentence-level labels for facet--sentence alignment.}
{For each facet $f$, we further prompt the LLM with the query paper $d_q$ and its facet-positive papers $\mathcal{D}_{\mathrm{pos}}^{(f)}$ to generate one scientific question $q^{(f)}$ grounded in the supplied content and associated with the target facet. The prompt templates are provided in Appendix~\ref{sec:prompts}.}

{The annotated papers and generated questions are organized into query-centered FTUs:}
\begin{equation}
\label{eq:ftu}
\text{FTU} = \{d_q,\mathcal{D}_\textit{pos}^{(f)},\mathcal{D}_\textit{neg}^{(f)},q^{(f)} \mid f \in \{bg,mt,rs\}\}.
\end{equation}
{Each FTU combines facet-specific paper relations, sentence-level labels, and three generated questions under the same facet scheme, supporting joint contrastive learning and facet--sentence alignment.} {An example from the constructed FTUs is provided in Appendix~\ref{sec:ftu-example}.}

\subsubsection{{Data Quality Assessment.}}
\label{sec:data-quality}
{We conduct a manual quality assessment of sentence-level
facet annotations and generated questions. We randomly sample 100 distinct papers containing a total of 821 abstract sentences from the metadata of the constructed FTU corpus, along with 300 generated questions equally distributed across the three facets. Two annotators independently evaluate the sampled data according to the following criteria.} 

{For sentence-level annotations, the annotators assign each sentence to background, method, or result based on its primary semantic function within the complete title-and-abstract context, without viewing the LLM-generated labels. For generated questions, the annotators make separate binary judgments on two criteria:
(1) facet alignment, assessing whether the primary information need matches the target facet; and
(2) clarity and grounding, assessing whether the information need is clearly expressed and its key concepts and premises are supported by the papers supplied during generation.} 

{Table~\ref{tab:ftu-quality} summarizes the assessment results. For each criterion, agreement is computed as the proportion of items receiving identical judgments from the two annotators before adjudication. Disagreements are then resolved through discussion to establish final reference labels and acceptance judgments. A sentence annotation is accepted when its LLM-generated label matches the adjudicated reference label, while a question is accepted when it satisfies the corresponding criterion. The high acceptance of sentence labels (97.69\%) indicates that the LLM-generated annotations generally agree with human interpretations of the three semantic roles. Strong question--facet alignment (99.33\%) suggests that the generator follows the explicit target-facet requirements in the prompts. Clarity and grounding receive a lower acceptance rate, highlighting that following a facet instruction does not automatically ensure a clearly formulated and fully source-supported information need. This criterion requires both coherent expression and content fidelity, making it a complementary check on question quality. The high agreement between independent annotators further supports the consistent interpretation and application of the evaluation criteria.}

\begin{table}[h!]
\centering
\caption{{Manual quality assessment of FTUs.
Accepted counts and rates are based on adjudicated
judgments, while agreement is the percentage of identical
judgments between the two annotators before adjudication.}}
\label{tab:ftu-quality}
\footnotesize
\setlength{\tabcolsep}{8pt}
\renewcommand{\arraystretch}{0.9}
\begin{tabular}{lccc}
\toprule
{Criterion} & {Accepted / Evaluated} & {Rate} & {Agreement} \\
\midrule
{Sentence facet labeling}
& {802 / 821}
& {97.69\%}
& {96.83\%} \\

{Question facet alignment}
& {298 / 300}
& {99.33\%}
& {96.67\%} \\

{Question clarity and grounding}
& {285 / 300}
& {95.00\%}
& {92.67\%} \\
\bottomrule
\end{tabular}
\end{table}

\subsubsection{{Data Statistics and Splits.}}
\label{data statistics}
{The constructed corpus contains 216,078 FTUs, covering 7,428,874 distinct facet-labeled papers and 648,234 generated questions, with three questions per unit. Table~\ref{tab:data} summarizes their distribution across 19 scientific fields, with Medicine, Computer Science, and Biology accounting for the largest shares.} {Field assignments are not mutually exclusive, where items associated with multiple fields are counted once in each corresponding field. Paper and question percentages are calculated separately over their respective summed field counts.}
{The distribution exhibits a long-tail pattern
across scientific fields, while providing diverse scientific terminology, research contexts, and methodological content, offering supervision for learning shared representations across heterogeneous scientific topics.}

{We divide the corpus at the unit level, allocating 205,274 units (95\%) for training and 10,804 units (5\%) for development. All paper relations and generated questions belonging to the same FTU remain in the same split, preserving the grouped supervision associated with each query paper. The development split is used for architectural design and hyperparameter selection in Section~\ref{exp:architecture}.}

\begin{table}[t]
\centering
\caption{{Scientific-field distribution of the constructed
FTU corpus.
Papers include query papers and positive/negative
papers; query papers are the subset serving as FTU centers, each generating three facet-specific questions.
The percentage columns correspond to papers and questions,
respectively.}}
\label{tab:data}
\footnotesize
\setlength{\tabcolsep}{8pt}
\renewcommand{\arraystretch}{0.95}
\begin{tabular}{lccccc}
\toprule
{Scientific Field}
& {\# Papers}
& {Percentage}
& {\# Query Papers}
& {\# Questions}
& {Percentage} \\
\midrule
Medicine
& 3,002,908 & {34.19\%}
& 101,065 & {303,195} & {32.71\%} \\

Computer Science
& 1,538,273 & {17.51\%}
& 77,079 & {231,237} & {24.95\%} \\

Biology
& 1,158,449 & {13.19\%}
& 38,820 & {116,460} & {12.56\%} \\

Psychology
& 494,617 & {5.63\%}
& 14,017 & {42,051} & {4.54\%} \\

Chemistry
& 400,837 & {4.56\%}
& 10,400 & {31,200} & {3.37\%} \\

Physics
& 384,276 & {4.37\%}
& 9,904 & {29,712} & {3.21\%} \\

Mathematics
& 371,410 & {4.23\%}
& 16,852 & {50,556} & {5.45\%} \\

Engineering
& 310,521 & {3.54\%}
& 10,280 & {30,840} & {3.33\%} \\

Materials Science
& 251,316 & {2.86\%}
& 7,624 & {22,872} & {2.47\%} \\

Economics
& 201,989 & {2.30\%}
& 8,253 & {24,759} & {2.67\%} \\

Business
& 168,263 & {1.92\%}
& 4,328 & {12,984} & {1.40\%} \\

Environmental Science
& 107,915 & {1.23\%}
& 2,570 & {7,710} & {0.83\%} \\

Geology
& 100,061 & {1.14\%}
& 2,003 & {6,009} & {0.65\%} \\

Geography
& 93,067 & {1.06\%}
& 1,514 & {4,542} & {0.49\%} \\

Sociology
& 91,846 & {1.05\%}
& 1,604 & {4,812} & {0.52\%} \\

Political Science
& 74,960 & {0.85\%}
& 2,470 & {7,410} & {0.80\%} \\

Art
& 12,470 & {0.14\%}
& 59 & {177} & {0.02\%} \\

Philosophy
& 11,572 & {0.13\%}
& 24 & {72} & {0.01\%} \\

History
& 9,271 & {0.11\%}
& 109 & {327} & {0.04\%} \\
\bottomrule
\end{tabular}
\end{table}

\subsection{Training Procedure and Objectives}
Given the constructed Facet-Aware Training Units (FTUs), we next describe how they are processed to optimize UniFAR through facet-aware training objectives.

\begin{figure*}[!t]
  \centering
  \includegraphics[width=0.98\linewidth]{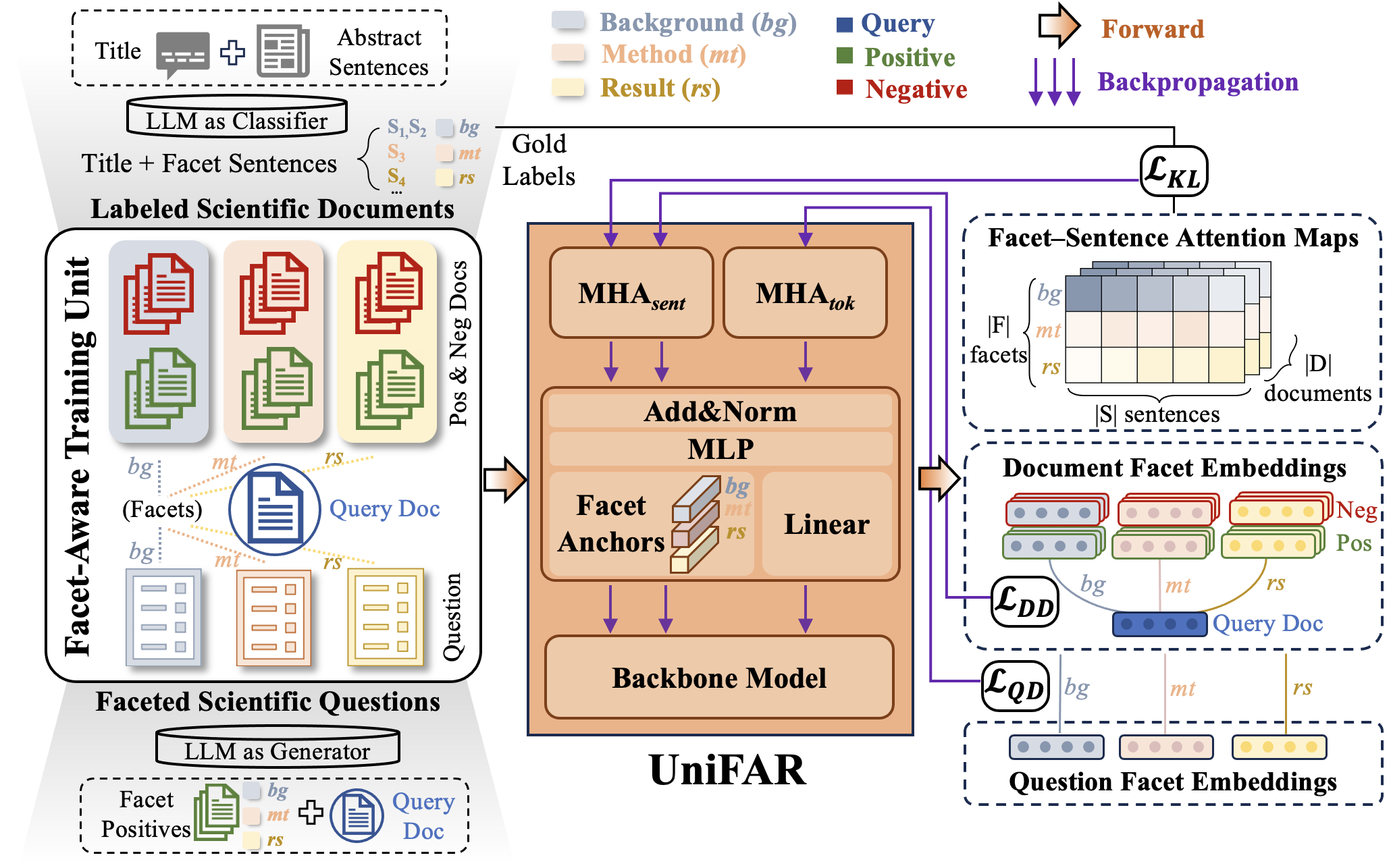}
  \caption{The facet-aware joint training strategy of UniFAR. Each \textbf{Facet-Aware Training Unit (FTU, left)} is constructed via LLM-based facet labeling and question generation, containing a query document, its facet-specific positives and negatives, and facet-aware scientific questions. FTUs are then forwarded through \textbf{UniFAR (middle)} as described in Section~\ref{sec:framework} to produce attention maps and facet-level document/question embeddings for \textbf{Joint Optimization (right)}. Three objectives: (1) $\mathcal{L}_\textit{DD}$ for doc–doc retrieval, (2) $\mathcal{L}_\textit{QD}$ for q–doc retrieval, and (3) $\mathcal{L}_\textit{KL}$ for facet-sentence attention alignment, are jointly optimized through backpropagation to update parameters of UniFAR.}
  \label{fig_2}
\end{figure*}

\subsubsection{Forward Computation}
Each FTU is fed into UniFAR for forward computation, where all inputs are encoded by the replaceable backbone encoder (Section~\ref{encode}) and then passed through the facet aggregation module (Section~\ref{aggregation}). 
Specifically, the query document $d_q$, facet-specific positive and negative documents $\mathcal{D}^{(f)}_{\text{pos}}$ and $\mathcal{D}^{(f)}_{\text{neg}}$, as well as facet-aware questions $q^{(f)}$, are independently encoded into contextual representations and subsequently aggregated into facet-level embeddings.

Through facet-aware attention, the model produces a set of embeddings $\{\mathbf{e}^{(f)}\}_{f\in\mathcal{F}}$ for each input, capturing different semantic aspects such as background, method, and result. 
Meanwhile, the aggregation process yields a facet--sentence attention matrix that reflects how each facet selectively attends to relevant sentences within the document.
As illustrated on the right of Figure~\ref{fig_2}, these embeddings and attention maps are then used to compute contrastive objectives and supervise facet-level alignment in subsequent optimization.

\subsubsection{Facet-Aware Contrastive Objectives}
To enable effective multi-positive and multi-negative learning under facet-aware supervision, we adopt the InfoNCE loss \citep{InfoNCE} as the contrastive learning objective, instead of the triplet margin loss \cite{margin} used in prior methods \cite{specter,SciNCL,FLeW,FaBle}. 
Given an anchor embedding $\mathbf{a}$, a set of positive embeddings $\mathcal{P} = \{\mathbf{p}_i\}$  and a set of negative embeddings $\mathcal{N} = \{\mathbf{n}_m\}$, the loss function is defined as:
\begin{equation}
    \mathcal{L}_{\text{CL}}
    =
    -\frac{1}{|\mathcal{P}|}
    \sum\limits_{p_i\in\mathcal{P}}
    \log
    \frac{
    e^{\operatorname{sim}(\mathbf{a},\mathbf{p}_i)/\tau}
    }{
    \sum\limits_{p_j\in\mathcal{P}} e^{\operatorname{sim}(\mathbf{a},\mathbf{p}_j)/\tau}
    +
    \sum\limits_{n_m\in\mathcal{N}} e^{\operatorname{sim}(\mathbf{a},\mathbf{n}_m)/\tau}
    },
\end{equation}
where $\mathrm{sim}(\cdot)$ denotes a similarity function and $\tau$ is the temperature hyperparameter.

To capture both document-centric similarity and question-driven relevance, we apply the contrastive objective to both doc-doc and q-doc retrieval under each facet. 
Specifically, for each facet $f$, we define two complementary objectives:
\begin{equation}
  \mathcal{L}^{(f)}_{\text{DD}}
  =
  \mathcal{L}_{\text{CL}}
  \big(
    \mathbf{a}=\mathbf{d}_q,\;
    \mathcal{P}=\{\mathbf{d}^{(f)}_{\textit{+}}\},\;
    \mathcal{N}=\{\mathbf{d}^{(f)}_{-}\}
  \big)
  ,
\label{eq:dd}
\end{equation}
\begin{equation}
  \mathcal{L}^{(f)}_{\text{QD}}
  =
  \mathcal{L}_{\text{CL}}
  \big(
    \mathbf{a}=\mathbf{q}^{(f)},\;
    \mathcal{P}=\{\mathbf{d}_q\}\cup\{\mathbf{d}^{(f)}_{\textit{+}}\},\;
    \mathcal{N}=\{\mathbf{d}^{(f)}_{-}\}
  \big),
\label{eq:qd}
\end{equation}
where {\small$\mathbf{d}_q$} is the query document embedding, {\small$\mathbf{d}^{(f)}_{\textit{+}}$} and {\small$\mathbf{d}^{(f)}_{-}$} are the facet-specific positive and negative document embeddings, and {\small$\mathbf{q}^{(f)}$} is the facet-aware question embedding.
The doc-doc objective $\mathcal{L}^{(f)}_{\text{DD}}$ preserves citation-based semantic similarity within each facet, while the q-doc objective $\mathcal{L}^{(f)}_{\text{QD}}$ aligns natural-language questions with relevant documents. 
Together, they enable a unified embedding space that supports both retrieval settings through shared facet-aware representations.

\subsubsection{Facet-Level Supervision via Attention Alignment}
To explicitly supervise facet-aware aggregation beyond implicit contrastive signals, we introduce a KL-divergence loss that aligns the predicted facet–sentence attention with sentence-level facet annotations. 
For each document $d$, let $\mathcal{S}^{(d)} = \{s_1,\ldots,s_{|S^{(d)}|}\}$ denote its sentence set, and $\mathcal{F}=\{bg,mt,rs\}$ denote the facet set. UniFAR produces a facet–sentence attention matrix 
$\mathbf{A}^{(d)} \in \mathbb{R}^{|\mathcal{F}|\times |\mathcal{S}^{(d)}|}$ through the facet aggregation module, where each entry is defined as:
\begin{equation}
A^{(d)}_{f,s_i} \;=\; \mathrm{Attn}(q_f, s_i),
\label{eq:attention_matrix}
\end{equation}
where $q_f$ is the input-specific query for facet $f$, and $s_i$ is the representation of sentence $i$. Each row $\mathbf{A}^{(d)}_{f,:}$ captures the attention distribution of facet $f$ over all sentences.

To construct the gold alignment, we leverage sentence-level facet labels obtained via LLM-based classification. Each sentence $s_i$ is assigned a facet label $y_i \in \mathcal{F}$. Based on these labels, we define the gold facet–sentence matrix $\mathbf{G}^{(d)} \in \mathbb{R}^{|\mathcal{F}|\times |\mathcal{S}^{(d)}|}$ as:
\begin{equation}
G^{(d)}_{f,s_i} \;=\; \mathbb{I}(y_i = f),
\label{eq:gold_matrix}
\end{equation}
where $\mathbb{I}(\cdot)$ is the indicator function. Thus, each column of $\mathbf{G}^{(d)}$ contains a one-hot assignment indicating the facet membership of sentence $s_i$.

To make $\mathbf{A}^{(d)}$ and $\mathbf{G}^{(d)}$ comparable, we normalize both matrices along the sentence dimension (row-wise):
\begin{equation}
\tilde{A}^{(d)}_{f,s_i} = \frac{A^{(d)}_{f,s_i}}{\sum_{j} A^{(d)}_{f,s_j}}, 
\quad
\tilde{G}^{(d)}_{f,s_i} = \frac{G^{(d)}_{f,s_i}}{\sum_{j} G^{(d)}_{f,s_j}}.
\label{eq:norm}
\end{equation}

This normalization transforms each row into a probability distribution over sentences, ensuring that the alignment is performed at the facet level. The KL-alignment loss is then defined as:
\begin{equation}
  \mathcal{L}_{\text{KL}}^{(d)} = \frac{1}{|\mathcal{F}^{(d)}|} \sum\limits_{f \in \mathcal{F}^{(d)}} \frac{1}{|\mathcal{S}^{(d)}|} \sum\limits_{s_i \in \mathcal{S}^{(d)}} \tilde{G}^{(d)}_{f,s_i} \log\!\frac{\tilde{G}^{(d)}_{f,s_i}}{\tilde{A}^{(d)}_{f,s_i}}.
\label{eq:kl}
\end{equation}

\subsubsection{Overall Objective and Training Procedure}
The overall training objective integrates the facet-level contrastive losses and the document-level KL-alignment loss as follows:
\begin{equation}
  \mathcal{L}_{\text{total}} = \frac{1}{|\mathcal{F}|} \sum\limits_{f\in\mathcal{F}}
  \Big(
    \mathcal{L}_{\text{DD}}^{(f)}
    +
    \mathcal{L}_{\text{QD}}^{(f)}
  \Big)
  + \lambda \frac{1}{|\mathcal{D}|} \sum\limits_{d\in\mathcal{D}} \mathcal{L}_{\mathrm{KL}}^{(d)},
\label{eq:total_loss}
\end{equation}
where $\mathcal{F}=\{\textit{background},\textit{method},\textit{result}\}$ and $\lambda$ is a weighting coefficient. 
The complete training procedure is summarized in Algorithm~\ref{alg:training}. 

\begin{algorithm}[!h]
  \captionsetup{font=small}
  \caption{Facet-Aware Joint Training Procedure of UniFAR}
  \label{alg:training}
  \footnotesize
  \begin{algorithmic}[1]
  \REQUIRE Training set $\mathcal{T}$ of FTUs; facet set $\mathcal{F}=\{bg,mt,rs\}$; temperature $\tau$; weight $\lambda$
  \ENSURE Parameters $\Theta$ of UniFAR
  
  \STATE Initialize $\Theta$
  \FOR{mini-batch $\mathcal{B} \subset \mathcal{T}$}
      \STATE $\mathcal{L}_{DD} \leftarrow 0$, $\mathcal{L}_{QD} \leftarrow 0$, $\mathcal{L}_{KL} \leftarrow 0$
      \STATE $\mathcal{D}_{\mathcal{B}} \leftarrow \emptyset$
  
      \FOR{each FTU $u \in \mathcal{B}$}
      \STATE Decompose $u$ as $\{d_q,\mathcal{D}^{(f)}_{\mathrm{pos}},\mathcal{D}^{(f)}_{\mathrm{neg}},q^{(f)}\}_{f\in\mathcal{F}}$
      \STATE $\mathcal{D}_u \leftarrow \{d_q\}\cup \bigcup_{f\in\mathcal{F}}\left(\mathcal{D}^{(f)}_{\mathrm{pos}} \cup \mathcal{D}^{(f)}_{\mathrm{neg}}\right)$
      \STATE $\mathcal{D}_{\mathcal{B}} \leftarrow \mathcal{D}_{\mathcal{B}} \cup \mathcal{D}_u$

      \STATE Encode all inputs $\{d_q,\mathcal{D}^{(f)}_{\mathrm{pos}},\mathcal{D}^{(f)}_{\mathrm{neg}},q^{(f)}\}$ with the backbone model \quad (Eq.~\ref{eq:token}, \ref{eq:sentence})
      \STATE Apply facet fusion $\rightarrow \{\mathbf{e}_{x}^{(f)}\}_{f\in\mathcal{F}}$ for each input $x$, and $\mathbf{A}^{(d)}$ for each document $d$ \quad (Eq.~\ref{eq:attention}, \ref{eq:fusion}, \ref{eq:attention_matrix})
      \FOR{each facet $f \in \mathcal{F}$}
        \STATE $\mathcal{L}_{DD} \leftarrow \mathcal{L}_{DD} + \mathcal{L}^{(f)}_{DD}$ \quad (Eq.~\ref{eq:dd})
        \STATE $\mathcal{L}_{QD} \leftarrow \mathcal{L}_{QD} + \mathcal{L}^{(f)}_{QD}$ \quad (Eq.~\ref{eq:qd})
      \ENDFOR

      \FOR{each document $d \in \mathcal{D}_u$}
          \STATE Construct gold matrix $\mathbf{G}^{(d)}$ from sentence-level facet labels \quad (Eq.~\ref{eq:gold_matrix})
          \STATE Normalize $\mathbf{A}^{(d)}, \mathbf{G}^{(d)} \rightarrow \tilde{\mathbf{A}}^{(d)}, \tilde{\mathbf{G}}^{(d)}$ \quad (Eq.~\ref{eq:norm})
          \STATE $\mathcal{L}_{KL} \leftarrow \mathcal{L}_{KL} + \mathcal{L}^{(d)}_{KL}$ \quad (Eq.~\ref{eq:kl})
      \ENDFOR
    \ENDFOR

    \STATE $\mathcal{L}_{\mathrm{total}} \leftarrow
    \frac{1}{|\mathcal{F}|}\left(\mathcal{L}_{DD}+\mathcal{L}_{QD}\right)
    + \lambda \frac{1}{|\mathcal{D}_{\mathcal{B}}|}\mathcal{L}_{KL}$ \quad (Eq.~\ref{eq:total_loss})

    \STATE {Backpropagate $\mathcal{L}_{total}$ to Update $\Theta$}
  \ENDFOR
\end{algorithmic}
\end{algorithm}

\section{Experiments}
\label{sec:experiments}
In this section, we comprehensively evaluate UniFAR through the following research questions:
\begin{itemize}[label=---]
  \item {\textit{RQ1}: How does UniFAR compare with baselines in q-doc
  and doc-doc retrieval benchmarks from different fields and across backbone encoders?}
  \item {\textit{RQ2}: How does UniFAR compare with instruction-following models in facet-based retrieval?}
  \item {\textit{RQ3.} How do different facet matching strategies affect retrieval performance?}
  \item {\textit{RQ4}: How do UniFAR's key components contribute to retrieval performance?}
  \item {\textit{RQ5}}: How does UniFAR capture semantic facets across scientific questions and documents?
  \item {\textit{RQ6}: What are UniFAR's efficiency trade-offs relative
  to its backbones and other baselines?}
\end{itemize}

\subsection{Experimental Setup}
\subsubsection{Benchmark Datasets.} 
\label{benchmarks}
{We conduct evaluations on five benchmarks covering q-doc and doc-doc
retrieval across computer science, medicine, and biomedicine.} {To assess generalization across retrieval
tasks and scientific fields, we use the configuration determined
on the FTU development split (Section~\ref{exp:architecture})
without further tuning on any benchmark.} {Table~\ref{tab:benchmarks} summarizes the statistics of these benchmarks.} 

\begin{itemize}[label=---]
\item {\textit{DORIS-MAE} \citep{dorismae} evaluates q-doc retrieval
for complex scientific information needs. Its 100 human-authored questions
are decomposed into aspects. Evaluating the full questions, individual
aspects, and combinations of aspects yields 1,203 retrieval instances.}

\item {\textit{LitSearch} \citep{litsearch} evaluates q-doc retrieval
for broad and specific questions about machine learning and natural
language processing papers. Questions are either generated from
inline citation contexts or written by authors about their own papers.}

\item {\textit{CSFCube} \citep{csfcube} evaluates doc-doc retrieval
given a seed paper and a target facet (background, method, or result).
Graded relevance judgments indicate how closely each candidate
matches the seed paper on the specified facet.}

\item {\textit{TREC-COVID} \citep{treccovid} evaluates q-doc retrieval
for 50 COVID-19 research queries. Each is expressed in three
forms: a keyword query, a natural-language question, and a narrative
description. We evaluate each form separately using the same relevance
judgments.}

\item {\textit{RELISH} \citep{relish} evaluates doc-doc retrieval
in biomedicine. Each query is a seed paper, and experts provide graded relevance judgments of candidate papers to the seed.}

\end{itemize}

\begin{table*}[h!]
  \centering
  \caption{{Statistics of the five retrieval benchmarks.}
}
  \label{tab:benchmarks}
  \footnotesize
  \setlength{\tabcolsep}{4pt}
  \renewcommand{\arraystretch}{0.95}
  \begin{tabular}{@{}cccccccc@{}}
    \toprule
    Benchmark & {Task} & {Domain}
    & \multicolumn{3}{c}{{Query Breakdown}}
    & {\# Queries} & \# Candidates \\
    \midrule

    \multirow{2}{*}{DORIS-MAE}
    & \multirow{2}{*}{{Q-Doc}}
    & {Computer} & \textit{Query} & \textit{Aspect} & \textit{Subquery\_2}
    & \multirow{2}{*}{{100}} & \multirow{2}{*}{363,133} \\
    & & {Science} & 100 & 100 & 1,003 & & \\
    \midrule

    \multirow{3}{*}{LitSearch}
    & \multirow{3}{*}{{Q-Doc}}
    & \multirow{3}{*}{{\shortstack[c]{Computer\\Science}}}
    & & \textit{Broad} & \textit{Specific}
    & \multirow{3}{*}{597} & \multirow{3}{*}{64,183} \\
    & & & \textit{Inline} & 120 & 231 & & \\
    & & & \textit{Author} & 35  & 211 & & \\
    \midrule

    \multirow{2}{*}{CSFCube}
    & \multirow{2}{*}{{Doc-Doc}}
    & {Computer} & \textit{Background} & \textit{Method} & \textit{Result}
    & \multirow{2}{*}{50} & \multirow{2}{*}{4,207} \\
    & & {Science} & 16 & 17 & 17 & & \\
    \midrule

    \multirow{2}{*}{{TREC-COVID}}
    & \multirow{2}{*}{{Q-Doc}}
    & \multirow{2}{*}{{Medicine}}
    & {\textit{Keyword}} & {\textit{Question}} & {\textit{Narrative}}
    & \multirow{2}{*}{{50}} & \multirow{2}{*}{{171,332}} \\
    & & & {50} & {50} & {50} & & \\
    \midrule
    {RELISH} & {Doc-Doc} & {Biomedicine}
    & \multicolumn{3}{c}{{---}} & {3,190} & {191,245} \\
    \bottomrule
  \end{tabular}
\end{table*}

\subsubsection{Evaluation Metrics.} 
For \textit{DORIS-MAE}, we report Recall@5, R-Precision, and MAP.
For \textit{CSFCube}, we report MAP and nDCG$_{\%20}$, with nDCG
computed over the top 20\% of each query's ranked candidate list
\citep{aspire,csfcube}. For \textit{LitSearch}, we report Recall@$k$,
MAP@$k$, and nDCG@$k$ at $k \in \{5,10,20\}$.
{For \textit{TREC-COVID},
we report nDCG@10 and MAP separately for the keyword, question, and narrative
inputs. For \textit{RELISH}, we compute AP
and nDCG within each seed paper's candidate pool, and report MAP and
mean nDCG across seed papers.}
{Metric definitions are provided in Appendix~\ref{app:metrics}.} {To assess the statistical significance of UniFAR’s improvements over the best baseline on the benchmark metrics, we use a two-sided paired t-test.}

\subsubsection{{Baseline Models and Backbone Encoders.}}
\label{baselines}
{We evaluate UniFAR from two perspectives: its performance relative
to established scientific retrievers and the effect of applying it
to different backbone encoders.}

\emph{Scientific retrieval baselines.}
{SciBERT \citep{scibert} is pretrained on scientific corpora and has
been widely used as a backbone for scientific retrievers. We select representative baselines that build upon SciBERT and its derivatives:
(1) \textit{Document-level representation methods} encode each paper into a single dense vector, including SPECTER \citep{specter}, SciNCL \citep{SciNCL}, and SPECTER-2 \citep{specter2}; 
(2) \textit{Fine-grained or facet-aware methods} incorporate sentence-level or facet-level semantic signals during representation learning or matching, including CoSentBERT \citep{aspire}, Aspire \citep{aspire}, FaBle \citep{FaBle}, and FLeW \citep{FLeW}. Aspire has three variants: \textit{{\small TS}} (text-supervised single-match), \textit{{\small OT}} (optimal transport-based multi-match), and \textit{{\small TS+OT}}, which combines both objectives.}

\emph{Replaceable backbone encoders.}
{We use SciBERT as the primary backbone for comparing UniFAR with the
scientific baselines described above. Since our framework
supports replaceable backbone encoders, we also train and evaluate
UniFAR with three general-purpose retrievers: Contriever \citep{Contriever},}
{BGE-base \citep{BGE}, and E5-base \citep{E5}.} {For each backbone, we use its name alone to denote the original encoder
and the prefixes \textit{FTU-} and \textit{UniFAR-} to distinguish two
fine-tuned variants. Both variants start from the same backbone checkpoint
and use the same FTUs and contrastive objectives.
\textit{\textbf{FTU-{\footnotesize\{Backbone\}}}} represents each input with a single vector, without facet
anchors, KL alignment loss, or multi-vector matching.
\textit{\textbf{UniFAR-{\footnotesize\{Backbone\}}}} learns three facet embeddings through
anchor-based aggregation and facet--sentence supervision, and uses
these embeddings for facet-aware matching.
These comparisons distinguish the effect of training on FTU data from the contribution of the facet-aware framework.}

\subsection{{Configuration and Implementation}}
\subsubsection{{Training Settings.}}
{We use the same settings for all FTU-based training runs
across backbones above, including the development experiments below.}
{We train each model for two
epochs using AdamW with linear warm-up over the first 5\% of training
steps. Training uses four NVIDIA RTX 3090 GPUs (24\,GB each), with a
per-device batch size of 4 and four gradient accumulation steps,
yielding an effective batch size of 16 per GPU.
The contrastive temperature $\tau$ is set to 0.08. 
We adopt a lower learning rate of $2\times10^{-5}$ for the backbone model and shared facet anchors to preserve pretrained knowledge, and a higher rate of
$5\times10^{-5}$ for the other aggregation modules to facilitate
faster adaptation.}

\subsubsection{{Design and Hyperparameter Selection.}}
\label{exp:architecture}
{With these optimization settings fixed, we use SciBERT as the
reference backbone to compare architectural variants and select the
facet--sentence alignment loss weight $\lambda$ on the FTUs development
split (FTU-dev). 
For each facet of each development FTU, Doc--Doc (DD) uses the seed paper to
rank the facet-specific positive and negative papers, and Q--Doc (QD) uses the
generated question to rank the same candidates plus the seed paper,
which is also treated as relevant. We compute nDCG and AP over the candidate papers for each FTU--facet pair and average over all such pairs to obtain
nDCG and MAP. 
The reference configuration uses keyword-based anchor initialization
with orthogonalization. For weight fusion, it uses a sentence-count threshold
of $\gamma=3$ to select the sentence-level branch when the input contains at least 3 sentences. It jointly optimizes DD and QD with a KL loss weight
of $\lambda=0.5$. Gradients from $\mathcal{L}_{\mathrm{QD}}$ are
stopped on the document side, and gradients from
$\mathcal{L}_{\mathrm{KL}}$ are stopped at the backbone encoder.
Table~\ref{tab:ftu-dev-design} compares this configuration with
variants grouped by design factor on FTU-dev.}

\begin{table*}[t]
\centering
\caption{{Design and hyperparameter comparisons on FTU-dev with SciBERT backbone.
Variants are grouped by design factor, with other settings held fixed.
Bold indicates the retained reference configuration.
Orth.\ denotes orthogonalization, $\gamma$ is the hard threshold of sentence-counting for branch selection, and $w$ is the
weight on the sentence-level branch.}}
\label{tab:ftu-dev-design}
\footnotesize
\setlength{\tabcolsep}{8pt}
\renewcommand{\arraystretch}{0.9}
\begin{tabular}{@{}cclccccc@{}}
\toprule
\multirow{2}{*}{{Design Factor}}
& \multicolumn{2}{c}{\multirow{2}{*}{{Configuration}}}
& \multicolumn{2}{c}{{Doc-Doc (DD)}}
& \multicolumn{2}{c}{{Q-Doc (QD)}}
& \multirow[c]{2}{*}[-0.5ex]{{\shortstack{Avg. nDCG}}} \\
\cmidrule(lr){4-5}\cmidrule(lr){6-7}
& \multicolumn{2}{l}{}
& {nDCG} & {MAP} & {nDCG} & {MAP} & \\
\midrule

{--}
& \multicolumn{2}{c}{{\textbf{Reference}}}
& {\textbf{0.8959}} & {\textbf{0.7689}} & {\textbf{0.9264}} & {\textbf{0.8066}} & {\textbf{0.9112}} \\
\midrule

\multirow{3}{*}{{\shortstack[c]{Anchor\\Initialization}}}
& \multicolumn{2}{c}{{\textit{Random, no Orth.}}}
& {0.8775} & {0.7501} & {0.9066} & {0.7787} & {0.8921} \\
& \multicolumn{2}{c}{{\textit{Random + Orth.}}}
& {0.8812} & {0.7519} & {0.9087} & {0.7832} & {0.8950} \\
& \multicolumn{2}{c}{{\textit{Keywords, no Orth.}}}
& {0.8913} & {0.7622} & {0.9245} & {0.8009} & {0.9079} \\
\midrule

\multirow{2}{*}{{\shortstack[c]{Gradient\\Propagation}}}
& \multicolumn{2}{c}{{\textit{No document detach for} $\mathcal{L}_{\mathrm{QD}}$}}
& {0.8841} & {0.7570} & {0.9217} & {0.8012} & {0.9029} \\
& \multicolumn{2}{c}{{\textit{No backbone detach for} $\mathcal{L}_{\mathrm{KL}}$}}
& {0.8754} & {0.7494} & {0.8979} & {0.7705} & {0.8867} \\
\midrule

\multirow{10}{*}{{\shortstack[c]{Weight Fusion\\Strategy}}}
& \multirow{4}{*}{{\shortstack[c]{\textit{Branch}\\\textit{selection}}}}
& {$\gamma=1$} & {0.8945} & {0.7677} & {0.9216} & {0.7921} & {0.9081} \\
& & {$\gamma=2$} & {0.8948} & {0.7676} & {0.9251} & {0.8039} & {0.9100} \\
& & {$\gamma=4$} & {0.8952} & {0.7681} & {0.9261} & {0.8059} & {0.9107} \\
& & {$\gamma=5$} & {0.8956} & {0.7686} & {0.9253} & {0.8041} & {0.9105} \\
\cmidrule(l){2-8}
& \multirow{5}{*}{{\shortstack[c]{\textit{Weighted}\\\textit{interpolation}}}}
& {$w=0$} & {0.8921} & {0.7631} & {0.9255} & {0.8049} & {0.9088} \\
& & {$w=0.25$} & {0.8928} & {0.7633} & {0.9250} & {0.8036} & {0.9089} \\
& & {$w=0.5$}  & {0.8936} & {0.7648} & {0.9238} & {0.8025} & {0.9087} \\
& & {$w=0.75$} & {0.8938} & {0.7645} & {0.9231} & {0.7965} & {0.9085} \\
& & {$w=1$}    & {0.8945} & {0.7677} & {0.9216} & {0.7921} & {0.9081} \\
\cmidrule(l){2-8}
& \multicolumn{2}{c}{{\textit{Learnable soft gating}}}
& {0.8964} & {0.7698} & {0.9261} & {0.8077} & {0.9113} \\
\midrule

\multirow{4}{*}{{KL Weight}}
& \multicolumn{2}{c}{{$\lambda=0$}}
& {0.8798} & {0.7549} & {0.9103} & {0.7933} & {0.8951} \\
& \multicolumn{2}{c}{{$\lambda=0.25$}}
& {0.8874} & {0.7616} & {0.9209} & {0.8008} & {0.9042} \\
& \multicolumn{2}{c}{{$\lambda=0.75$}}
& {0.8937} & {0.7644} & {0.9232} & {0.8025} & {0.9085} \\
& \multicolumn{2}{c}{{$\lambda=1.0$}}
& {0.8940} & {0.7651} & {0.9245} & {0.8059} & {0.9093} \\
\bottomrule
\end{tabular}
\end{table*}

{\emph{(1) Anchor initialization.}
For each facet, we encode representative seed keywords with the corresponding backbone encoder (SciBERT in this experiment)
and average their embeddings, followed by normalization and
orthogonalization to initialize the facet anchors.} {Keyword
initialization provides a semantic starting point associated with
background, method, or result, while orthogonalization separates
the initial anchor directions. Orthogonality is imposed only at initialization, while the anchors remain
trainable and can learn relationships across facets during
subsequent optimization. Keyword initialization outperforms random initialization, and
orthogonalization further improves performance. These results support combining semantic initialization
with distinct initial directions, helping the anchors learn
representations associated with different facets.}

{\emph{(2) Gradient propagation.}
Document embeddings are detached when computing
$\mathcal{L}_{\mathrm{QD}}$ to limit the direct influence of generated
questions on the document branch and reduce potential conflicts
between DD and QD updates on that branch.} {Removing this operation
lowers average nDCG from 0.9112 to 0.9029, with declines on both
tasks.} {Backbone outputs are also detached for
$\mathcal{L}_{\mathrm{KL}}$, allowing sentence-level labels to guide
how the facet module organizes semantic information, while the backbone remains optimized by the contrastive objectives. This separation aims to preserve a stable
semantic foundation for facet aggregation.} {Propagating KL gradients
into the backbone lowers average nDCG to 0.8867, supporting the
restriction of alignment supervision to the facet module.}

{\emph{(3) Weight fusion strategy.}
Branch selection uses the sentence-count threshold $\gamma$ to
choose between token- and sentence-level attention. This allows
short inputs with few sentence units ($|S|<\gamma$) to use finer-grained token
information, while longer inputs use sentence-level context ($|S|\geq\gamma$).}
{The threshold $\gamma=3$ achieves the highest average nDCG among
the tested thresholds (0.9112), exceeding the best fixed
interpolation (0.9089). Learnable soft gating reaches 0.9113,
providing only a small additional gain. For fixed interpolation, we use the same weight across facets,
i.e., $w_f=w$ for all $f\in\mathcal{F}$. Hard routing evaluates
one MHA branch per input, whereas interpolation with $0<w<1$
and soft gating evaluate both branches before fusion. It therefore
avoids computing the unselected branch and requires no learned
gating parameters. Considering both retrieval performance and
branch computation, we retain hard routing with $\gamma=3$.}

{\emph{(4) KL weight.}
The weight $\lambda$ balances facet--sentence alignment against
the contrastive objectives. Removing alignment supervision
($\lambda=0$) lowers average nDCG from 0.9112 to 0.8951, indicating
that sentence labels help the model learn facet representations
useful for retrieval. Increasing $\lambda$ beyond 0.5 also reduces
performance, suggesting that stronger alignment supervision does
not necessarily produce better retrieval representations.
We retain $\lambda=0.5$, which achieves the best DD and QD
performance among the tested weights and provides a suitable
balance between semantic facet alignment and retrieval learning.}

{Based on these comparisons, we retain the reference configuration,
which balances retrieval performance with the computational efficiency
of single-branch aggregation. We use this configuration for all
subsequent benchmark evaluations across backbone encoders, without
further benchmark-specific tuning.}

\subsection{{Main Results (RQ1)}}

\begin{table*}[!t]
  \centering
  \footnotesize
  \setlength{\tabcolsep}{3pt}
  \renewcommand{\arraystretch}{0.8}
  \begin{threeparttable}
  \caption{{Performance comparison on the DORIS-MAE benchmark (q-doc) across different query levels. The best results within each backbone group are highlighted in bold.}}
  \label{tab1}
  \begin{tabular}{@{}cccccccccc@{}}
  \toprule
  Level & \multicolumn{3}{c}{Query} & \multicolumn{3}{c}{Aspect} & \multicolumn{3}{c}{Subquery\_2} \\
  \cmidrule(l){2-4} \cmidrule(l){5-7} \cmidrule(l){8-10}
  Model / Metric & {Recall@5} & {MAP} & {R-Precision} & {Recall@5} & {MAP} & {R-Precision} & {Recall@5} & {MAP} & {R-Precision} \\
  \midrule
  SciBERT & 0.0513 & 0.2034 & 0.1713 & 0.0541 & 0.2166 & 0.1850 & 0.0243 & 0.2123 & 0.1756 \\
  SPECTER & 0.1334 & 0.3575 & 0.3266 & 0.1156 & 0.3139 & 0.2843 & 0.1051 & 0.3827 & 0.3540 \\
  SciNCL & 0.1321 & 0.3610 & 0.3138 & 0.1269 & 0.3543 & 0.3075 & 0.1075 & 0.4748 & 0.4371 \\
  SPECTER-2 & 0.1450 & 0.3712 & 0.3341 & 0.1297 & 0.3577 & {0.3130} & 0.1114 & 0.4726 & 0.4373 \\
  {\scriptsize TS}Aspire & 0.1426 & 0.3700 & 0.3381 & 0.1430 & 0.3731 & 0.3363 & \textbf{0.1315} & 0.4585 & 0.4146 \\
  {\scriptsize OT}Aspire & 0.1334 & 0.3670 & 0.3363 & 0.1445 & 0.3609 & 0.3253 & 0.1256 & 0.4553 & 0.4060 \\
  FLeW & 0.1442 & 0.3872 & 0.3271 & 0.1424 & 0.3756 & 0.3328 & 0.1143 & 0.4766 & 0.4401 \\
  {\textit{FTU{\tiny -SciBERT}}} & {0.1430} & {0.3779} & {0.3317} & {0.1318} & {0.3625} & {0.3196} & {0.1102} & {0.4669} & {0.4287} \\
  \textit{UniFAR{\tiny -SciBERT}} & \textbf{0.1504$^{**}$} & \textbf{0.3903$^{*}$} & \textbf{0.3401$^{*}$} & \textbf{0.1458$^{**}$} & \textbf{0.3782$^{*}$} & \textbf{0.3386$^{*}$} & 0.1208 & \textbf{0.4801$^{*}$} & \textbf{0.4425$^{*}$} \\
  \midrule
  Contriever & 0.1190 & 0.3344 & 0.2870 & 0.1115 & 0.2991 & 0.2657 & 0.0949 & 0.4307 & 0.4015 \\
  {\textit{FTU{\tiny -Contriever}}} & {0.1454} & {0.3503} & {0.3165} & {0.1204} & {0.3493} & {0.3093} & {0.1052} & {0.4655} & {0.4322} \\
  \textit{UniFAR{\tiny -Contriever}} & \textbf{0.1606{$^{**}$}} & \textbf{0.3752{$^{**}$}} & \textbf{0.3398{$^{**}$}} & \textbf{0.1375{$^{**}$}} & \textbf{0.3541{$^{**}$}} & \textbf{0.3139{$^{*}$}} & \textbf{0.1135{$^{**}$}} & \textbf{0.4738{$^{*}$}} & \textbf{0.4410{$^{**}$}} \\
  \midrule
  {BGE-base} & {0.1811} & {0.4209} & {0.3768} & {0.1827} & {0.4187} & {0.3726} & {0.1336} & {0.5143} & {0.4705} \\
  {\textit{FTU{\tiny -BGE}}} & {0.1853} & {0.4234} & {0.3779} & {0.1845} & {0.4234} & {0.3729} & {0.1332} & {0.5166} & {0.4725} \\
  {\textit{UniFAR{\tiny -BGE}}} & {\textbf{0.1887$^{*}$}} & {\textbf{0.4302$^{*}$}} & {\textbf{0.3815$^{*}$}} & {\textbf{0.1912$^{**}$}} & {\textbf{0.4298$^{**}$}} & {\textbf{0.3775$^{*}$}} & {\textbf{0.1378$^{**}$}} & {\textbf{0.5200$^{*}$}} & {\textbf{0.4774$^{**}$}} \\
  \midrule
  {E5-base} & {0.1551} & {0.3879} & {0.3429} & {0.1498} & {0.3781} & {0.3395} & {0.1267} & {0.5091} & {0.4676} \\
  {\textit{FTU{\tiny -E5}}} & {0.1598} & {0.3902} & {0.3480} & {0.1534} & {0.3823} & {0.3412} & {0.1285} & {0.5098} & {0.4707} \\
  {\textit{UniFAR{\tiny -E5}}} & {\textbf{0.1611}} & {\textbf{0.3943$^{*}$}} & {\textbf{0.3521$^{*}$}} & {\textbf{0.1587$^{*}$}} & {\textbf{0.3865$^{*}$}} & {\textbf{0.3420}} & {\textbf{0.1324$^{**}$}} & {\textbf{0.5137$^{*}$}} & {\textbf{0.4752$^{**}$}} \\
  \bottomrule
  \end{tabular}
  \begin{tablenotes}
    \footnotesize
    \item * and ** denote statistical significance compared to the best baseline with $p < 0.1$ and $p < 0.05$, respectively.
  \end{tablenotes}
\end{threeparttable}
\end{table*}

\begin{table*}[h!]
  \footnotesize
  \centering
  \setlength{\tabcolsep}{4.5pt}
  \renewcommand{\arraystretch}{0.8}
  \begin{threeparttable}
  \caption{{Performance comparison on the LitSearch benchmark (q–doc) across three metrics, reported at cutoff levels of 5, 10, and 20. The best results within each backbone group are highlighted in bold.}}
  \label{tab:litsearch}
  \begin{tabular}{@{}cccccccccc@{}}
    \toprule
    Metric & \multicolumn{3}{c}{Recall} & \multicolumn{3}{c}{MAP} & \multicolumn{3}{c}{nDCG} \\
    \cmidrule(lr){2-4} \cmidrule(lr){5-7} \cmidrule(l){8-10}
    Model & @5 & @10 & @20 & @5 & @10 & @20 & @5 & @10 & @20 \\
    \midrule
    SPECTER & 0.2278 & 0.3015 & 0.3685 & 0.1548 & 0.1640 & 0.1687 & 0.1727 & 0.1955 & 0.2124 \\
    SciNCL & 0.2881 & 0.3702 & 0.4405 & 0.1978 & 0.2086 & 0.2136 & 0.2199 & 0.2459 & 0.2636 \\
    SPECTER-2 & 0.3467 & 0.4422 & 0.5209 & 0.2262 & 0.2397 & 0.2450 & 0.2553 & 0.2870 & 0.3068 \\
    {\scriptsize TS}Aspire & 0.2178 & 0.2797 & 0.3585 & 0.1440 & 0.1523 & 0.1578 & 0.1614 & 0.1814 & 0.2011 \\
    {\scriptsize OT}Aspire & 0.2211 & 0.2764 & 0.3568 & 0.1502 & 0.1611 & 0.1689 & 0.1735 & 0.1944 & 0.2058 \\
    FLeW & 0.3417 & 0.4472 & 0.5243 & 0.2261 & 0.2353 & 0.2459 & 0.2536 & 0.2853 & 0.3073 \\
    {\textit{FTU{\tiny -SciBERT}}} & {0.3501} & {0.4344} & {0.5278} & {0.2302} & {0.2411} & {0.2482} & {0.2530} & {0.2903} & {0.3093} \\
    \textit{UniFAR{\tiny -SciBERT}} & \textbf{0.3518$^{*}$} & \textbf{0.4523$^{*}$} & \textbf{0.5461$^{**}$} & \textbf{0.2449$^{**}$} & \textbf{0.2571$^{**}$} & \textbf{0.2627$^{**}$} & \textbf{0.2709$^{**}$} & \textbf{0.2959$^{*}$} & \textbf{0.3204$^{**}$} \\
    \midrule
  Contriever & 0.2429 & 0.3049 & 0.3735 & 0.1603 & 0.1682 & 0.1726 & 0.1798 & 0.1995 & 0.2163 \\
  {\textit{FTU{\tiny -Contriever}}} & {0.3084} & {0.3555} & {0.4275} & {0.2071} & {0.2177} & {0.2230} & {0.2161} & {0.2326} & {0.2714} \\
  \textit{UniFAR{\tiny -Contriever}} & \textbf{0.3300{$^{**}$}} & \textbf{0.4020{$^{**}$}} & \textbf{0.4757{$^{**}$}} & \textbf{0.2386{$^{**}$}} & \textbf{0.2482{$^{**}$}} & \textbf{0.2534{$^{**}$}} & \textbf{0.2602{$^{**}$}} & \textbf{0.2834{$^{**}$}} & \textbf{0.3022{$^{**}$}} \\
  \midrule
  {BGE-base} & {0.5109} & {0.5930} & {0.6750} & {0.3873} & {0.3989} & {0.4048} & {0.4169} & {0.4440} & {0.4653} \\
  {\textit{FTU{\tiny -BGE}}} & {0.5212} & {0.6025} & {0.6853} & {0.3915} & {0.4035} & {0.4124} & {0.4264} & {0.4557} & {0.4748} \\
  {\textit{UniFAR{\tiny -BGE}}} & {\textbf{0.5345$^{**}$}} & {\textbf{0.6133$^{**}$}} & {\textbf{0.6972$^{**}$}} & {\textbf{0.4028$^{**}$}} & {\textbf{0.4116$^{**}$}} & {\textbf{0.4206$^{**}$}} & {\textbf{0.4368$^{**}$}} & {\textbf{0.4680$^{**}$}} & {\textbf{0.4839$^{**}$}} \\
  \midrule
  {E5-base} & {0.4573} & {0.5544} & {0.6214} & {0.3359} & {0.3484} & {0.3535} & {0.3655} & {0.3961} & {0.4137} \\
  {\textit{FTU{\tiny -E5}}} & {0.4685} & {0.5662} & {0.6321} & {0.3423} & {0.3556} & {0.3610} & {0.3786} & {0.4065} & {0.4312} \\
  {\textit{UniFAR{\tiny -E5}}} & {\textbf{0.4760$^{*}$}} & {\textbf{0.5748$^{**}$}} & {\textbf{0.6437$^{**}$}} & {\textbf{0.3542$^{**}$}} & {\textbf{0.3672$^{**}$}} & {\textbf{0.3723$^{**}$}} & {\textbf{0.3903$^{**}$}} & {\textbf{0.4132$^{**}$}} & {\textbf{0.4555$^{**}$}} \\
    \bottomrule
  \end{tabular}
  \begin{tablenotes}
    \footnotesize
    \item * and ** denote statistical significance compared to the best baseline with $p < 0.1$ and $p < 0.05$, respectively.
  \end{tablenotes}
\end{threeparttable}
\end{table*}

{We evaluate the overall effectiveness of UniFAR on the five benchmarks introduced in Section~\ref{benchmarks}, covering q-doc and doc--doc retrieval tasks across computer science, medicine, and biomedicine. The framework supports replaceable backbones, so we evaluate UniFAR on the four encoders described in Section~\ref{baselines}.} {For each
backbone, we compare the original encoder with its single-vector \textit{FTU}
and multi-vector \textit{UniFAR} variants.} {In the SciBERT group, we also compare
UniFAR with established scientific retrievers, including both single-vector and fine-grained methods.}

\subsubsection{{Multi-Level Q-Doc Retrieval.}}
{On DORIS-MAE (Table~\ref{tab1}), \textit{UniFAR{\scriptsize -SciBERT}} leads its group
on all metrics at the \textit{Query} and \textit{Aspect} levels.
A full query can express several information needs, whereas an aspect
focuses on one part of the question. The gains at both levels suggest
that facet-aware matching is useful when the scope of relevance changes
from the complete question to a specific requirement. At the
\textit{Subquery\_2} level, {\scriptsize TS}Aspire achieves higher
Recall@5 (0.1315 vs.\ 0.1208), while UniFAR achieves higher MAP
(0.4801 vs.\ 0.4585) and R-Precision. This distinction matters for
queries that combine aspects. Sentence-level matching remains
competitive for retrieving relevant papers in the first few positions,
while UniFAR achieves better retrieval quality as measured across the
relevant results.}

\subsubsection{{Realistic Q-Doc Retrieval.}}
{\textit{UniFAR{\scriptsize -SciBERT}} leads its group across all metrics and cutoffs
on LitSearch (Table~\ref{tab:litsearch}). Its gains are visible both
near the top of the ranking and at greater retrieval depth. It
improves nDCG@5 over SPECTER-2 (0.2709 vs.\ 0.2553) and retrieves
a larger share of relevant papers in the top 20 than FLeW
(Recall@20 of 0.5461 vs.\ 0.5243). The combination is useful for
literature search, where users may inspect a few results first and
then expand their search to cover more relevant work. Both Aspire
variants trail several single-vector scientific retrievers on this
benchmark. Their performance suggests that representing a paper at
a finer granularity is only part of the problem. The matching
function also needs to identify which aspects of that paper address
the question.}

\begin{table*}[t]
  \centering
  \footnotesize
  \setlength{\tabcolsep}{5.5pt}
  \renewcommand{\arraystretch}{0.8}
  \begin{threeparttable}
  \caption{{Performance comparison on the CSFCube benchmark (doc-doc) across three facets and their aggregated scores. The best results within each backbone group are highlighted in bold.}}
  \label{tab2}
  \begin{tabular}{@{}ccccccccc@{}}
  \toprule
  Facet & \multicolumn{2}{c}{Background} & \multicolumn{2}{c}{Method} & \multicolumn{2}{c}{Result} & \multicolumn{2}{c}{Aggregated} \\
  \cmidrule(l){2-3} \cmidrule(l){4-5} \cmidrule(l){6-7} \cmidrule(l){8-9}
  Model / Metric & MAP & nDCG{\%20} & MAP & nDCG{\%20} & MAP & nDCG{\%20} & MAP & nDCG{\%20} \\
  \midrule
  SciBERT & 0.2589 & 0.4499 & 0.1923 & 0.3441 & 0.2306 & 0.4206 & 0.2263 & 0.4033 \\
  CoSentBERT & 0.3578 & 0.6127 & 0.1927 & 0.3877 & 0.3215 & 0.5068 & 0.2895 & 0.5068 \\
  SPECTER & 0.4395 & 0.6670 & 0.2244 & 0.3741 & 0.3679 & 0.5667 & 0.3423 & 0.5328 \\
  SciNCL & 0.4964 & 0.7002 & 0.2714 & 0.4661 & 0.4183 & 0.6170 & 0.3937 & 0.5924 \\
  SPECTER-2 & 0.5079 & 0.7092 & \textbf{0.2719} & 0.4728 & 0.3965 & 0.6373 & 0.3986 & 0.6042 \\
  {\scriptsize TS+OT}Aspire & 0.5179 & 0.7099 & 0.2668 & 0.4760 & 0.4306 & 0.6482 & 0.4026 & 0.6086 \\
  FaBle\tiny{-SPECTER} & 0.4266 & 0.6738 & 0.2598 & 0.4497 & 0.3860 & 0.5810 & 0.3560 & 0.5660 \\
  FLeW\scriptsize{-bg/mt/rs} & 0.5093 & 0.7109 & 0.2376 & 0.4437 & 0.3903 & 0.6237 & 0.3801 & 0.5998 \\
  {\textit{FTU{\tiny -SciBERT}}} & {0.5158} & {0.7158} & {0.2451} & {0.4521} & {0.4258} & {0.6393} & {0.3996} & {0.6035} \\
  \textit{UniFAR{\tiny -SciBERT}} & \textbf{0.5191$^{*}$} & \textbf{0.7255$^{**}$} & 0.2634 & \textbf{0.4772$^{*}$} & \textbf{0.4319$^{*}$} & \textbf{0.6488} & \textbf{0.4033} & \textbf{0.6127$^{**}$} \\
  \midrule
  Contriever & 0.4317 & 0.6564 & 0.2311 & 0.4135 & 0.3845 & 0.6241 & 0.3474 & 0.5625 \\
  {\textit{FTU{\tiny -Contriever}}} & {0.4509} & {0.6643} & {0.2390} & {0.4347} & {0.3866} & {0.6300} & {0.3527} & {0.5727} \\
  \textit{UniFAR{\tiny -Contriever}} & \textbf{0.4729{$^{**}$}} & \textbf{0.6990{$^{**}$}} & \textbf{0.2471{$^{**}$}} & \textbf{0.4506{$^{**}$}} & \textbf{0.3896} & \textbf{0.6344{$^{*}$}} & \textbf{0.3683{$^{**}$}} & \textbf{0.5881{$^{**}$}} \\
  \midrule
  {BGE-base} & {0.5132} & {0.7189} & {0.2706} & {0.4857} & {0.3972} & {0.6270} & {0.3948} & {0.6028} \\
  {\textit{FTU{\tiny -BGE}}} & {0.5165} & {0.7274} & {0.2819} & {0.4927} & {0.4153} & {0.6365} & {0.4011} & {0.6113} \\
  {\textit{UniFAR{\tiny -BGE}}} & {\textbf{0.5221$^{**}$}} & {\textbf{0.7349$^{*}$}} & {\textbf{0.2886$^{**}$}} & {\textbf{0.4982$^{*}$}} & {\textbf{0.4210$^{*}$}} & {\textbf{0.6422$^{*}$}} & {\textbf{0.4035}} & {\textbf{0.6192$^{*}$}} \\
  \midrule
  {E5-base} & {0.5237} & {0.7280} & {0.2601} & {0.4751} & {0.3838} & {0.6015} & {0.3848} & {0.5989} \\
  {\textit{FTU{\tiny -E5}}} & {0.5315} & {0.7324} & {0.2689} & {0.4823} & {0.3909} & {0.6135} & {0.3965} & {0.6074} \\
  {\textit{UniFAR{\tiny -E5}}} & {\textbf{0.5386$^{**}$}} & {\textbf{0.7402$^{*}$}} & {\textbf{0.2755$^{**}$}} & {\textbf{0.4916$^{**}$}} & {\textbf{0.4024$^{**}$}} & {\textbf{0.6322$^{**}$}} & {\textbf{0.4023$^{**}$}} & {\textbf{0.6168$^{**}$}} \\
  \bottomrule
  \end{tabular}
  \begin{tablenotes}
    \footnotesize
    \item * and ** denote statistical significance compared to the best baseline with $p < 0.1$ and $p < 0.05$, respectively.
    \item {\tiny -SPECTER} denotes models trained with SPECTER initialization.
    \item {\scriptsize -bg/mt/rs} indicates three facet encoders of FLeW, trained separately for background, method, and result.
  \end{tablenotes}
  \end{threeparttable}
\end{table*}

\subsubsection{{Faceted Doc-Doc Retrieval.}}
{CSFCube evaluates whether two papers are related with respect to a
specified facet (Table~\ref{tab2}). \textit{UniFAR{\scriptsize -SciBERT}} leads
its group on both aggregated metrics and on both metrics for
\textit{Background} and \textit{Result}. These results
suggest that the facet representations preserve distinctions useful
for comparing papers by their research context and by their findings.
The improvement across these facets is particularly useful in
query-by-example retrieval, where the seed paper alone leaves open
which aspect of similarity the user wants to explore. The \textit{Method} facet shows a more nuanced result. UniFAR leads
in nDCG at the reported cutoff, while SPECTER-2 retains higher MAP
(0.2719 vs.\ 0.2634).  The nDCG result indicates an advantage in
ordering papers by graded relevance within the evaluated portion of
the pools, while the MAP result gives SPECTER-2 the advantage in
average precision for this facet. UniFAR's lead on both aggregated
metrics reflects its performance across the facet-specific retrieval
tasks.}

\subsubsection{{Retrieval in Medicine and Biomedicine.}}
{Table~\ref{tab4} reports q-doc retrieval in medicine on TREC-COVID
and doc-doc retrieval in biomedicine on RELISH. On TREC-COVID, \textit{UniFAR{\scriptsize -SciBERT}} leads its
group for keywords, questions, and narratives. Full questions
yield its highest nDCG@10, while the largest nDCG@10 gain over
SPECTER-2 occurs for keywords (0.2631 vs.\ 0.2528). Thus, the
benefit extends to brief inputs that provide less explicit
context about the information need. RELISH tests a different
form of relevance by asking the model to rank biomedical papers
with respect to a seed paper. Here, \textit{UniFAR{\scriptsize -SciBERT}} improves MAP over
SPECTER-2 within the supplied candidate pools (0.8681 vs.\ 0.8612).
The gains on two benchmarks show that the effectiveness extends to medical and biomedical retrieval, where relevance is defined either by a textual information
need or by the relationship between papers.}
\begin{table*}[t]
  \centering
  \footnotesize
  \setlength{\tabcolsep}{5.5pt}
  \renewcommand{\arraystretch}{0.8}
  \begin{threeparttable}
  \caption{{Performance on the TREC-COVID (q-doc) and RELISH (doc-doc) benchmarks. The best results within each backbone group are highlighted in bold.}}
  \label{tab4}
  \begin{tabular}{@{}ccccccccc@{}}
    \toprule
  \multirow[c]{2}{*}[-0.5ex]{{Benchmark}}
  & \multicolumn{6}{c}{{TREC-COVID (q-doc)}}
  & \multicolumn{2}{c}{
      \multirow[c]{2}{*}[-0.5ex]{{\shortstack[c]{RELISH\\(doc-doc)}}}
    } \\
  \cmidrule{2-7}
  & \multicolumn{2}{c}{{Keyword}}
  & \multicolumn{2}{c}{{Question}}
  & \multicolumn{2}{c}{{Narrative}}
  & \multicolumn{2}{c}{} \\
  \cmidrule(lr){2-3}
  \cmidrule(lr){4-5}
  \cmidrule(lr){6-7}
  \cmidrule(l){8-9}
  {Model / Metric}
  & {nDCG@10} & {MAP}
  & {nDCG@10} & {MAP}
  & {nDCG@10} & {MAP}
  & {nDCG} & {MAP} \\
  \midrule
  {SciBERT} & {0.0508} & {0.0010} & {0.0417} & {0.0008} & {0.0205} & {0.0004} & {0.8280} & {0.7443} \\
  {SPECTER} & {0.1923} & {0.0211} & {0.3130} & {0.0590} & {0.3203} & {0.0655} & {0.9007} & {0.8410} \\
  {SciNCL} & {0.2375} & {0.0330} & {0.3607} & {0.0602} & {0.3278} & {0.0689} & {0.9067} & {0.8515} \\
  {SPECTER-2} & {0.2528} & {0.0386} & {0.3619} & {0.0809} & {0.3283} & {0.0740} & {0.9163} & {0.8612} \\
  {{\scriptsize TS}Aspire} & {0.1639} & {0.0132} & {0.3180} & {0.0563} & {0.3067} & {0.0723} & {0.8932} & {0.8297} \\
  {{\scriptsize OT}Aspire} & {0.1479} & {0.0275} & {0.3268} & {0.0636} & {0.3276} & {0.0729} & {0.8962} & {0.8355} \\
  {FLeW} & {0.1614} & {0.0095} & {0.3555} & {0.0450} & {0.3139} & {0.0568} & {0.9122} & {0.8583} \\
  {\textit{FTU{\tiny -SciBERT}}} & {0.2316} & {0.0258} & {0.3596} & {0.0678} & {0.3267} & {0.0635} & {0.9128} & {0.8597} \\
  {\textit{UniFAR{\tiny -SciBERT}}} & {\textbf{0.2631$^{**}$}} & {\textbf{0.0432$^{**}$}} & {\textbf{0.3672$^{**}$}} & {\textbf{0.0839$^{*}$}} & {\textbf{0.3329$^{*}$}} & {\textbf{0.0772$^{*}$}} & {\textbf{0.9213$^{**}$}} & {\textbf{0.8681$^{*}$}} \\
  \midrule
  {Contriever} & {0.1289} & {0.0051} & {0.1816} & {0.0101} & {0.1434} & {0.0110} & {0.8954} & {0.8286} \\
  {\textit{FTU{\tiny -Contriever}}} & {0.1830} & {0.0143} & {0.3179} & {0.0380} & {0.2398} & {0.0340} & {0.9042} & {0.8490} \\
  {\textit{UniFAR{\tiny -Contriever}}} & {\textbf{0.1970$^{**}$}} & {\textbf{0.0155$^{**}$}} & {\textbf{0.3411$^{**}$}} & {\textbf{0.0459$^{**}$}} & {\textbf{0.2427$^{**}$}} & {\textbf{0.0368$^{**}$}} & {\textbf{0.9110$^{**}$}} & {\textbf{0.8564$^{**}$}} \\
  \midrule
  {BGE-base} & {0.4196} & {0.1040} & {0.6306} & {0.2138} & {0.4785} & {0.1545} & {0.9164} & {0.8667} \\
  {\textit{FTU{\tiny -BGE}}} & {0.4251} & {0.1106} & {0.6425} & {0.2214} & {0.4876} & {0.1664} & {0.9189} & {0.8740} \\
  {\textit{UniFAR{\tiny -BGE}}} & {\textbf{0.4369$^{**}$}} & {\textbf{0.1187$^{**}$}} & {\textbf{0.6566$^{**}$}} & {\textbf{0.2337$^{**}$}} & {\textbf{0.5011$^{*}$}} & {\textbf{0.1752$^{**}$}} & {\textbf{0.9233$^{*}$}} & {\textbf{0.8793$^{*}$}} \\
  \midrule
  {E5-base} & {0.3417} & {0.1003} & {0.4854} & {0.1554} & {0.3099} & {0.0870} & {0.9113} & {0.8527} \\
  {\textit{FTU{\tiny -E5}}} & {0.3522} & {0.1075} & {0.5023} & {0.1620} & {0.3187} & {0.0934} & {0.9143} & {0.8609} \\
  {\textit{UniFAR{\tiny -E5}}} & {\textbf{0.3625$^{**}$}} & {\textbf{0.1096$^{*}$}} & {\textbf{0.5216$^{**}$}} & {\textbf{0.1688$^{*}$}} & {\textbf{0.3258$^{**}$}} & {\textbf{0.1002$^{**}$}} & {\textbf{0.9184$^{*}$}} & {\textbf{0.8712$^{**}$}} \\
  \bottomrule
  \end{tabular}
  \begin{tablenotes}
    \footnotesize
    \item {* and ** denote statistical significance compared to the best baseline with $p < 0.1$ and $p < 0.05$, respectively.}
  \end{tablenotes}
  \end{threeparttable}
\end{table*}

\subsubsection{{Performance Across Backbones.}}
{Across the four result tables (Tables 5$\sim$8),
UniFAR yields consistent improvements with SciBERT, Contriever,
BGE-base, and E5-base, demonstrating its applicability to both
scientific-domain and general-purpose encoders. In particular,
the gains with BGE and E5 show that facet-aware modeling remains
beneficial with strong pretrained retrievers. 
We organize
the comparisons within each backbone group to assess these benefits
from a common starting checkpoint, since absolute performance also
depends on the representation quality of the underlying encoder.}
{Beyond differences across backbones, we distinguish
the contributions of data and framework by comparing FTU-trained
single-vector variants (\textit{FTU-}) with the full facet-aware
variants (\textit{UniFAR-}). FTU training generally improves the
original encoders, and UniFAR further improves every reported
metric. For example, Contriever's LitSearch nDCG@10 rises from
0.1995 to 0.2326 with FTU training and to 0.2834 with UniFAR,
showing gains from both the data and the framework. This additional benefit also holds with a stronger
encoder and on doc-doc retrieval. For example,
\textit{UniFAR{\scriptsize -BGE}} improves RELISH MAP over
\textit{FTU{\scriptsize -BGE}} (0.8793 vs.\ 0.8740).
These comparisons highlight the complementary roles of the data
and the framework. FTUs provide facet-specific paper relations
and question-oriented supervision, while UniFAR preserves these
distinctions in separate embeddings and uses them during matching.
The gains over the matched FTU variants show that the facet-aware
framework further improves retrieval with the same training data.}

\subsection{{Instruction-based Retrieval (RQ2)}}
{We examine whether retrieval instructions can elicit facet-specific
embeddings from pretrained models and how these embeddings compare
with those learned by UniFAR on FTUs. We use GritLM-7B
\citep{gritlm}, an instruction-tuned model that supports both text
embedding and generation, and Qwen3Emb-8B \citep{qwen}, an embedding
model of comparable size that accepts task instructions. For each model, we compare
standard single-vector retrieval with an instructed multi-vector
variant that produces separate embeddings for background, method,
and result. We also evaluate UniFAR with Qwen3Emb-8B as its backbone,
providing a comparison between instructed and learned facet
representations with a shared starting encoder.}

\subsubsection{{Training Settings and Instructions.}}
{For UniFAR{\scriptsize -Qwen3Emb}, we freeze the unquantized BF16
Qwen3Emb-8B backbone and insert LoRA adapters into the query and
value projections (\texttt{q\_proj} and \texttt{v\_proj}) of all
36 self-attention layers. We set the LoRA rank $r=8$, $\alpha=16$, and dropout to $0.05$, and train the adapters jointly
with the facet module. Training uses four NVIDIA RTX 3090 GPUs (24\,GB each), with a
per-device batch size of 4 FTUs. With gradient checkpointing and gradient caching
in microbatches of four texts, the measured peak training memory
is approximately 20.98\,GB per GPU.}

{The single-vector and instructed facet variants of GritLM-7B and
Qwen3Emb-8B use their released weights. In the single-vector
setting, we prepend the benchmark's task instruction to the query
and encode candidates without a task instruction. In the facet
setting, we append the corresponding facet instruction to the
task instruction for each query. Each candidate receives only the facet instruction.
Table~\ref{tab:retrieval-instructions} lists the instructions,
which we fix before benchmark evaluation. We use each model's
native instruction template and pooling procedure. Each facet
embedding comes from a separate encoder pass.}

\subsubsection{{Performance.}} {Table~\ref{tab:instructed-retrieval} shows that facet instructions
are most useful when the benchmark specifies the facet to retrieve.
On CSFCube, the instructed variants improve \textit{Method}
retrieval for both encoders, and the instructed Qwen3Emb-8B
also improves on \textit{Result}. This suggests that a faceted instruction
can help the encoder emphasize a requested aspect of a paper.
The pattern changes when relevance is defined by the query or
seed paper as a whole. On LitSearch, TREC-COVID, and RELISH,
both instructed facet variants score below their single-vector
counterparts. The drop is particularly clear on TREC-COVID,
where Qwen3Emb-8B's score for natural-language questions falls
from 0.9143 to 0.8409. Encoding each facet separately can sharpen
the representation of an individual aspect, but the resulting
similarities still need to work together to rank papers for the
overall information need.}

{UniFAR{\scriptsize -Qwen3Emb} achieves the highest score in every
reported setting, including the CSFCube facets where instructed
embeddings are competitive. Starting from the same Qwen3Emb-8B
checkpoint, UniFAR learns how each facet should represent queries
and papers through FTU relevance supervision and facet--sentence
alignment. Its gains over both Qwen3Emb-8B variants suggest that
this shared supervision helps preserve distinctions among facets
while making their similarities useful for retrieval. On RELISH, UniFAR improves over the
strongest baseline (0.8487 vs.\ 0.8393), while its score on
TREC-COVID questions is close to that of the already strong
single-vector Qwen3Emb-8B. The clearest advantage is the
consistency of the learned facet representations across
facet-specific and general retrieval settings.}

\begin{table*}[t]
\centering
\footnotesize
\setlength{\tabcolsep}{5pt}
\renewcommand{\arraystretch}{1.1}
\caption{{Task and facet instructions used for the instructed
retrieval models.}}
\label{tab:retrieval-instructions}
\begin{tabular}{@{}cp{0.82\linewidth}@{}}
\toprule
\multicolumn{2}{c}{{\textbf{Task Instructions}}} \\
\midrule
{LitSearch}
& \raggedright {Given a research query, retrieve the title and abstract of the relevant research paper.} \tabularnewline
{CSFCube}
& \raggedright {Given a scientific paper query, retrieve relevant scientific paper abstracts.} \tabularnewline
{TREC-COVID}
& \raggedright {Given a scientific query, retrieve relevant documents.} \tabularnewline
{RELISH}
& \raggedright {Given a scientific paper, retrieve other relevant scientific papers.} \tabularnewline
\midrule
\multicolumn{2}{c}{{\textbf{Facet Instructions}}} \\
\midrule
{Background}
& \raggedright {Generate an embedding of the background facet of the following text, focusing on the research context, motivation, problem, and objectives.} \tabularnewline
{Method}
& \raggedright {Generate an embedding of the method facet of the following text, focusing on the proposed approach, model, techniques, and experimental procedure.} \tabularnewline
{Result}
& \raggedright {Generate an embedding of the result facet of the following text, focusing on the findings, outcomes, evidence, and conclusions.} \tabularnewline
\bottomrule
\end{tabular}
\end{table*}

\begin{table*}[t]
\centering
\footnotesize
\begin{threeparttable}
\caption{{Retrieval performance of instruction-based baselines and UniFAR evaluated by nDCG@10 scores across benchmarks.
The subscript \textit{facet} denotes variants that use a separate
instruction for each facet (background, method, and result) and use the
three resulting embeddings for facet-aware retrieval.}}
\label{tab:instructed-retrieval}
\setlength{\tabcolsep}{2.5pt}
\renewcommand{\arraystretch}{1.0}
\begin{tabular}{@{}cccccccccc@{}}
\toprule
\multirow{2}{*}{{Model / Benchmark}}
& \multirow[c]{2}{*}[-0.6ex]{{\shortstack[c]{LitSearch\\(q-doc)}}}
& \multicolumn{4}{c}{{CSFCube (doc-doc)}}
& \multicolumn{3}{c}{{TREC-COVID (q-doc)}}
& \multirow[c]{2}{*}[-0.6ex]{{\shortstack[c]{RELISH\\(doc-doc)}}} \\
\cmidrule(lr){3-6}\cmidrule(lr){7-9}
& & {{Background}} & {{Method}} & {{Result}}
  & {{Aggregated}} & {{Keyword}} & {{Question}}
  & {{Narrative}} & \\
\midrule
{GritLM-7B}
  & {0.5988} & {0.7298} & {0.4788} & {0.6369} & {0.6155}
  & {0.5528} & {0.7438} & {0.6658} & {0.8339} \\
{GritLM-7B\textsubscript{\scriptsize\itshape facet}}
  & {0.5859} & {0.7199} & {0.4868} & {0.6291} & {0.6124}
  & {0.4534} & {0.6703} & {0.6344} & {0.8279} \\
Qwen3Emb-8B
  & 0.6239 & {0.7414} & {0.4837} & {0.6502} & {0.6223}
  & {0.8450} & {0.9143} & {0.7891} & {0.8393} \\
{Qwen3Emb-8B\textsubscript{\scriptsize\itshape facet}}
  & {0.6164} & {0.7366} & {0.4949} & {0.6661} & {0.6252}
  & {0.7112} & {0.8409} & {0.6896} & {0.8239} \\
{UniFAR{\tiny -Qwen3Emb}}
  & \textbf{0.6311{$^{**}$}} & {\textbf{0.7566$^{**}$}}
  & {\textbf{0.4996$^{*}$}} & {\textbf{0.6772$^{**}$}}
  & {\textbf{0.6314$^{**}$}} & {\textbf{0.8511$^{*}$}}
  & {\textbf{0.9150}} & {\textbf{0.8023$^{**}$}}
  & {\textbf{0.8487$^{**}$}} \\
\bottomrule
\end{tabular}
\begin{tablenotes}
  \footnotesize
  \item {* and ** denote statistical significance compared to the best baseline with $p < 0.1$ and $p < 0.05$, respectively.}
\end{tablenotes}
\end{threeparttable}
\end{table*}

\subsection{{Facet Matching Strategies (RQ3)}}
\label{sec:matching}
{As described in Section~\ref{sec:facet_scoring}, the default strategy averages similarities between corresponding
facets. This preserves alignment between semantic roles and assesses
overall similarity across facets.} {However, evidence relevant to a query facet may appear under a different role in the candidate paper. Therefore, we compare and analyze more strategies to
examine how such cross-facet evidence affects retrieval.} 
{Let $\mathbf{M}_{q,c}^{(f,g)}$ denote the similarity between query
facet $f$ and candidate facet $g$, where
$f,g\in\mathcal{F}=\{\textit{bg},\textit{mt},\textit{rs}\}$.
The diagonal strategies compare each query facet with the corresponding
candidate facet. \textit{Diagonal Mean}, our
default strategy, gives these matches equal weight to assess overall
similarity. \textit{Diagonal Max} uses the strongest corresponding
match, allowing a single facet to determine the score even when
the other matches are weak.  The cross-facet strategies consider all
three candidate facets for each query facet and retain the highest
similarity in each row.  \textit{Cross-Facet Mean} averages
the three row maxima, so each query facet contributes its best match.
\textit{Cross-Facet Max} takes the largest row maximum, which is
equivalent to selecting the strongest match in the entire matrix.} 
\begin{equation*}
\begin{aligned}
\phi_{\mathrm{DiagMean}}
&= \tfrac{1}{3}\sum_{f\in\mathcal{F}}\mathbf{M}_{q,c}^{(f,f)},
& \phi_{\mathrm{DiagMax}}
&= \max_{f\in\mathcal{F}}\mathbf{M}_{q,c}^{(f,f)},\\
\phi_{\mathrm{CrossMean}}
&= \tfrac{1}{3}\sum_{f\in\mathcal{F}}
   \max_{g\in\mathcal{F}}\mathbf{M}_{q,c}^{(f,g)},
& \phi_{\mathrm{CrossMax}}
&= \max_{f,g\in\mathcal{F}}\mathbf{M}_{q,c}^{(f,g)}.
\end{aligned}
\end{equation*}

\subsubsection{{Strategy Performance.}}
{Table~\ref{tab:facet-matching} compares the four matching
strategies in terms of nDCG@10. On LitSearch, \textit{Cross-Facet Mean}
improves over \textit{Diagonal Mean} (0.3018 vs.\ 0.2959) and
outperforms \textit{Cross-Facet Max}. With mean aggregation held fixed, the gain suggests that cross-facet matching can help when the query and relevant evidence fall under different facets. For example, a question about a method may find relevant
evidence in the results of an experiment evaluating it. By averaging
the best match for each query facet, \textit{Cross-Facet Mean} lets
such evidence contribute alongside matches for the other facets.
Across all three TREC-COVID query formats and on RELISH,
\textit{Diagonal Mean} achieves the highest scores, and both
mean-based strategies outperform both max-based strategies. This
pattern suggests that combining evidence across facets contributes
more to retrieval on these benchmarks than expanding the scope of
facet matching. These results support \textit{Diagonal Mean} as a
broadly applicable default. When a task calls for matching evidence
across semantic roles, an alternative such as \textit{Cross-Facet
Mean} can yield further gains.}

\begin{table*}[t!]
\centering
\footnotesize
\caption{{Performance of four facet matching strategies evaluated
on UniFAR{\scriptsize -SciBERT} by nDCG@10 scores.}}
\label{tab:facet-matching}
\setlength{\tabcolsep}{7.5pt}
\renewcommand{\arraystretch}{1.0}
\begin{tabular}{@{}cccccc@{}}
\toprule
\multirow{2}{*}{{Strategy / Benchmark}}
& \multicolumn{3}{c}{{TREC-COVID (q-doc)}}
& \multirow[c]{2}{*}[-0.6ex]{{\shortstack[c]{RELISH\\(doc-doc)}}}
& \multirow[c]{2}{*}[-0.6ex]{{\shortstack[c]{LitSearch\\(q-doc)}}} \\
\cmidrule(lr){2-4}
& {Keyword} & {Question} & {Narrative} & & \\
\midrule
{\textit{Diagonal Mean}}
  & {\textbf{0.2631}} & {\textbf{0.3672}} & {\textbf{0.3329}} & {\textbf{0.7989}} & 0.2959 \\
{\textit{Diagonal Max}}
  & {0.2582} & {0.3635} & {0.3294} & {0.7688} & {0.2928} \\
{\textit{Cross-Facet Mean}}
  & {0.2615} & {0.3666} & {0.3300} & {0.7961} & {\textbf{0.3018}} \\
{\textit{Cross-Facet Max}}
  & {0.2527} & {0.3619} & {0.3244} & {0.7694} & {0.2961} \\
\bottomrule
\end{tabular}
\end{table*}

\subsubsection{{Case Analysis.}}
{On LitSearch, \textit{Cross-Facet Mean} achieves the highest nDCG@10
among the four strategies (0.3018 vs.\ 0.2959 for the default
\textit{Diagonal Mean}). The following case illustrates how allowing
cross-facet matches improves the rank of a relevant paper that the
default strategy places outside the top 10.
Table~\ref{tab:cross-facet-case-input} presents the full query and
the annotated relevant paper.}

{The query asks for a technique that reduces input tokens to speed up
inference in vision-language models. The abstract connects the
requested technique to its intended outcome. Sentences S0--S1 explain
the computational cost of cross-modal processing, S2--S3 describe how
pruning and merging reduce the token count, and S5 reports an increase
in inference throughput of up to twofold. The method description
identifies the technique sought by the query, and the reported result
provides empirical evidence that it delivers the requested efficiency
gain.} 

{The upper panel of Table~\ref{tab:litsearch-cross-facet-case} shows
why matching corresponding facets can miss evidence relevant to
the query. The query asks for a token-reduction technique with a
specific purpose, and the paper's measured gain shows
that the technique serves that purpose. This connection is reflected
in the stronger match between the query's \textit{Method} facet and
the paper's \textit{Result} facet (0.6278) than between the two
\textit{Method} facets (0.5591). The paper's \textit{Result} facet
is the strongest match for all three query facets. Under diagonal
matching, that evidence contributes only through the
\textit{Result}--\textit{Result} pair.}

{The lower panel of Table~\ref{tab:litsearch-cross-facet-case} shows the ranking consequence of different matching strategies. Under
\textit{Diagonal Mean}, the weaker \textit{Background} and
\textit{Method} matches lower the contribution of the strong
\textit{Result} match. \textit{Diagonal Max} brings the paper
into the top 10 by letting the strong corresponding match
determine its score. \textit{Cross-Facet Mean} gives each query
facet equal weight and allows each to match the paper's
\textit{Result} facet. With mean aggregation held fixed, this
change in matching scope moves the paper from 13th to second
within the same candidate pool.}

{\textit{Cross-Facet Max} assigns the paper the highest score among
the four strategies (0.6278), yet ranks it fourth (nDCG@10 of 0.4307).
Each strategy also rescores the other candidates, so a higher score
for the relevant paper does not mean a higher rank.
\textit{Cross-Facet Max} can assign a high score on the strength of
one facet pair alone, whereas \textit{Cross-Facet Mean} rewards high
similarities across the query facets. In this case, all three query
facets favor the same candidate facet, and averaging their best
matches produces the highest rank for the relevant paper.}

\begin{table*}[t]
\centering
\footnotesize
\caption{{Full query and annotated relevant paper for the LitSearch
case. Abstract sentences are indexed from S0 to S5 for reference.}}
\label{tab:cross-facet-case-input}
\begin{tabular}{@{}p{0.98\linewidth}@{}}
\toprule
\multicolumn{1}{c}{{\textbf{Query}}} \\
\midrule
{Which vision-language model paper in 2023 developed techniques
that reduce input tokens to improve model inference speed?} \\
\midrule
\multicolumn{1}{c}{{\textbf{Relevant Paper}}} \\
\midrule
{\textbf{Title:} PuMer: Pruning and Merging Tokens for Efficient
Vision Language Models} \\
{\textbf{Abstract:}
Large-scale vision language (VL) models use Transformers to perform
cross-modal interactions between the input text and image. (S0)}

{These cross-modal interactions are computationally expensive and
memory-intensive due to the quadratic complexity of processing the
input image and text. (S1)}

{We present PuMer: a token reduction framework that uses text-informed
Pruning and modality-aware Merging strategies to progressively reduce
the tokens of input image and text, improving model inference speed
and reducing memory footprint. (S2)}

{PuMer learns to keep salient image tokens related to the input text
and merges similar textual and visual tokens by adding lightweight
token reducer modules at several cross-modal layers in the VL model. (S3)}

{Training PuMer is mostly the same as finetuning the original VL model
but faster. (S4)}

{Our evaluation for two vision language models on four downstream VL
tasks shows PuMer increases inference throughput by up to 2x and
reduces memory footprint by over 50\% while incurring less than a
1\% accuracy drop. (S5)} \\
\bottomrule
\end{tabular}
\end{table*}

\begin{table}[t]
\centering
\footnotesize
\caption{{Facet similarities and retrieval ranks for the LitSearch case.
Rows and columns in the upper panel correspond to query and
candidate facets, respectively. The lower panel reports the
paper's score and rank and the query's nDCG@10 for each strategy
within the same candidate pool.}}
\label{tab:litsearch-cross-facet-case}
\begin{tabular}{@{}lccc@{}}
\toprule
{Query / Candidate} & {Background} & {Method} & {Result} \\
\midrule
{Background} & {0.4884} & {0.5678} & {0.6264} \\
{Method}     & {0.4723} & {0.5591} & {0.6278} \\
{Result}     & {0.4606} & {0.5448} & {0.6171} \\
\midrule
{Strategy} & {Score} & {Rank} & {nDCG@10} \\
\midrule
{\textit{Diagonal Mean}}    & {0.5549} & {13} & {0.0000} \\
{\textit{Diagonal Max}}     & {0.6171} &  {6} & {0.3562} \\
{\textit{Cross-Facet Mean}} & {0.6237} &  {2} & {0.6309} \\
{\textit{Cross-Facet Max}}  & {0.6278} &  {4} & {0.4307} \\
\bottomrule
\end{tabular}
\end{table}

\subsection{Ablation Studies ({RQ4})}
To address the misalignments between q-doc and doc-doc retrieval, UniFAR introduces three key components: multi-granularity aggregation for input granularity, facet semantics modeling for semantic complexity, and joint training objectives for supervision signals.
As shown in Table~\ref{tab3}, we conduct ablation studies to {validate the effectiveness and contribution of each component.}

\subsubsection{Multi-Granularity Aggregation.} Removing sentence-level attention (w/o $\text{MHA}_{sent}$) causes a larger drop on CSFCube (doc-doc), where scientific abstracts benefit from sentence-level semantic structure for facet aggregation. For example, the performance on the \textit{Result} facet decreases to 0.4163 (-3.6\%), and the \textit{Aggregated} score drops to 0.3876 (-3.9\%). Conversely, removing token-level attention (w/o $\text{MHA}_{tok}$) leads to greater degradation on DORIS-MAE (q-doc), where short queries depend more heavily on fine-grained attention over individual tokens for facet-relevant evidence. In particular, the \textit{Query} score decreases to 0.3796 (-2.7\%), and the \textit{Aspect} score drops to 0.3711 (-1.9\%). These degradation patterns indicate that combining sentence-level and token-level attention with adaptive granularity selection is essential for maintaining robust facet embeddings across different retrieval tasks.

\subsubsection{Facet Semantics Modeling.} Freezing the anchors (Freeze $\mathbf{A}_{facet}$) yields mild degradation across both benchmarks, indicating that static anchors initialized by facet words are still useful for structuring facet queries. For example, the \textit{Query} score decreases to 0.3806 (-2.5\%) on DORIS-MAE and \textit{Result} score to 0.4096 (-5.2\%) on CSFCube.  Removing the anchors (w/o $\mathbf{A}_{facet}$), the only structural source of facet-specific differentiation, causes all facet queries to collapse into the same vector, yielding identical facet embeddings. This collapse severely weakens facet disentanglement and leads to notable performance degradation on both benchmarks, with the \textit{Query} score dropping to 0.3693 (-5.4\%) and the \textit{Aspect} score to 0.3466 (-8.4\%) on DORIS-MAE. Removing the input-specific refinement (w/o $\mathbf{E}_{input}$) also reduces performance, e.g., the \textit{Result} score decreases to 0.4112 (-4.8\%) and the \textit{Background} score to 0.5083 (-2.1\%) on CSFCube, suggesting that anchors alone are insufficient and adapting them to each input is essential for capturing facet-specific nuances.

\subsubsection{Joint Training Objectives.} {To examine the benefit of jointly modeling doc-doc and q-doc
retrieval, we compare the full model with variants that retain
only one of the two contrastive objectives.
Joint training improves each task over its corresponding
single-task contrastive variant.} {On CSFCube, it outperforms
DD-only contrastive training (e.g., \textit{Aggregated}: 0.4033 vs.\ 0.3962).} {
On DORIS-MAE, it outperforms QD-only contrastive training
(e.g., \textit{Query}: 0.3903 vs.\ 0.3701), with the same trend across
the other query levels. These results suggest that inter-document
relations help learn paper representations useful for question-driven
retrieval, while question-oriented supervision also benefits
document matching. They provide empirical support for integrating
the two forms of supervision within a shared representation space.} Removing the KL-alignment loss (w/o $\mathcal{L}_{\textit{KL}}$) leads to significant performance drops on both benchmarks, indicating that facet–sentence attention alignment is crucial for effective facet-aware learning. For instance, the \textit{Aspect} score decreases to 0.3518 (-7.0\%) on DORIS-MAE, while the \textit{Method} score drops to 0.2375 (-9.8\%) on CSFCube. This supervisory signal provides consistent guidance for shaping facet-specific attention patterns and also enhances the interpretability of generated representations.

\begin{table*}[t]
  \centering
  \footnotesize
  \setlength{\tabcolsep}{4pt}
  \renewcommand{\arraystretch}{0.9}
  \caption{{Ablation studies on UniFAR{\scriptsize -SciBERT} evaluated by MAP on DORIS-MAE and CSFCube benchmarks.}}
  \label{tab3}
  \begin{tabular}{@{}ccccccccc@{}}
    \toprule
    \multicolumn{2}{c}{Benchmark}
    & \multicolumn{3}{c}{DORIS-MAE (q-doc)}
    & \multicolumn{4}{c}{CSFCube (doc-doc)} \\
    \cmidrule(lr){3-5}\cmidrule(l){6-9}
\multicolumn{2}{c}{{Component $\downarrow$ / Level $\rightarrow$}}
& Query & Aspect & Subquery\_2
& Background & Method & Result & Aggregated \\
    \midrule
    \multicolumn{2}{c}{\textbf{UniFAR (full)}}
    & \textbf{0.3903} & \textbf{0.3782} & \textbf{0.4801}
    & \textbf{0.5191} & \textbf{0.2634} & \textbf{0.4319} & \textbf{0.4033} \\
    \cmidrule(lr){1-9}
    \multirow[c]{2}{*}{{\shortstack[c]{Multi-Granularity\\Aggregation}}}
    & \textit{w/o $\text{MHA}_{sent}$}
    & 0.3890 & 0.3779 & 0.4769 & 0.5065 & 0.2523 & 0.4163 & 0.3876 \\
    & \textit{w/o $\text{MHA}_{tok}$}
    & 0.3796 & 0.3711 & 0.4695 & 0.5111 & 0.2476 & 0.4259 & 0.3894 \\
    \cmidrule(lr){1-9}
    \multirow[c]{3}{*}{{\shortstack[c]{Facet Semantics\\Modeling}}}
    & \textit{Freeze $\mathbf{A}_{facet}$}
    & 0.3806 & 0.3775 & 0.4788 & 0.5179 & 0.2537 & 0.4096 & 0.3913 \\
    & \textit{w/o $\mathbf{A}_{facet}$}
    & 0.3693 & 0.3466 & 0.4713 & 0.5142 & 0.2485 & 0.4166 & 0.3878 \\
    & \textit{w/o $\mathbf{E}_{input}$}
    & 0.3774 & 0.3737 & 0.4778 & 0.5083 & 0.2559 & 0.4112 & 0.3881 \\
    \cmidrule(lr){1-9}
    \multirow[c]{3}{*}{{\shortstack[c]{Joint Training\\Objectives}}}
    & {\textit{w/o $\mathcal{L}_{\textit{DD}}$}}
    & {0.3701} & {0.3525} & {0.4596} & {0.4876} & {0.2412} & {0.4184} & {0.3789} \\
    & \textit{w/o $\mathcal{L}_{\textit{QD}}$}
    & 0.3857 & 0.3766 & 0.4701 & 0.5159 & 0.2527 & 0.4308 & 0.3962 \\
    & \textit{w/o $\mathcal{L}_{\textit{KL}}$}
    & 0.3686 & 0.3518 & 0.4508 & 0.4941 & 0.2375 & 0.4226 & 0.3807 \\
    \bottomrule
  \end{tabular}
\end{table*}

\subsection{{Attention Visualization (RQ5)}}
\begin{figure*}[!t]
  \centering
  \includegraphics[width=\linewidth]{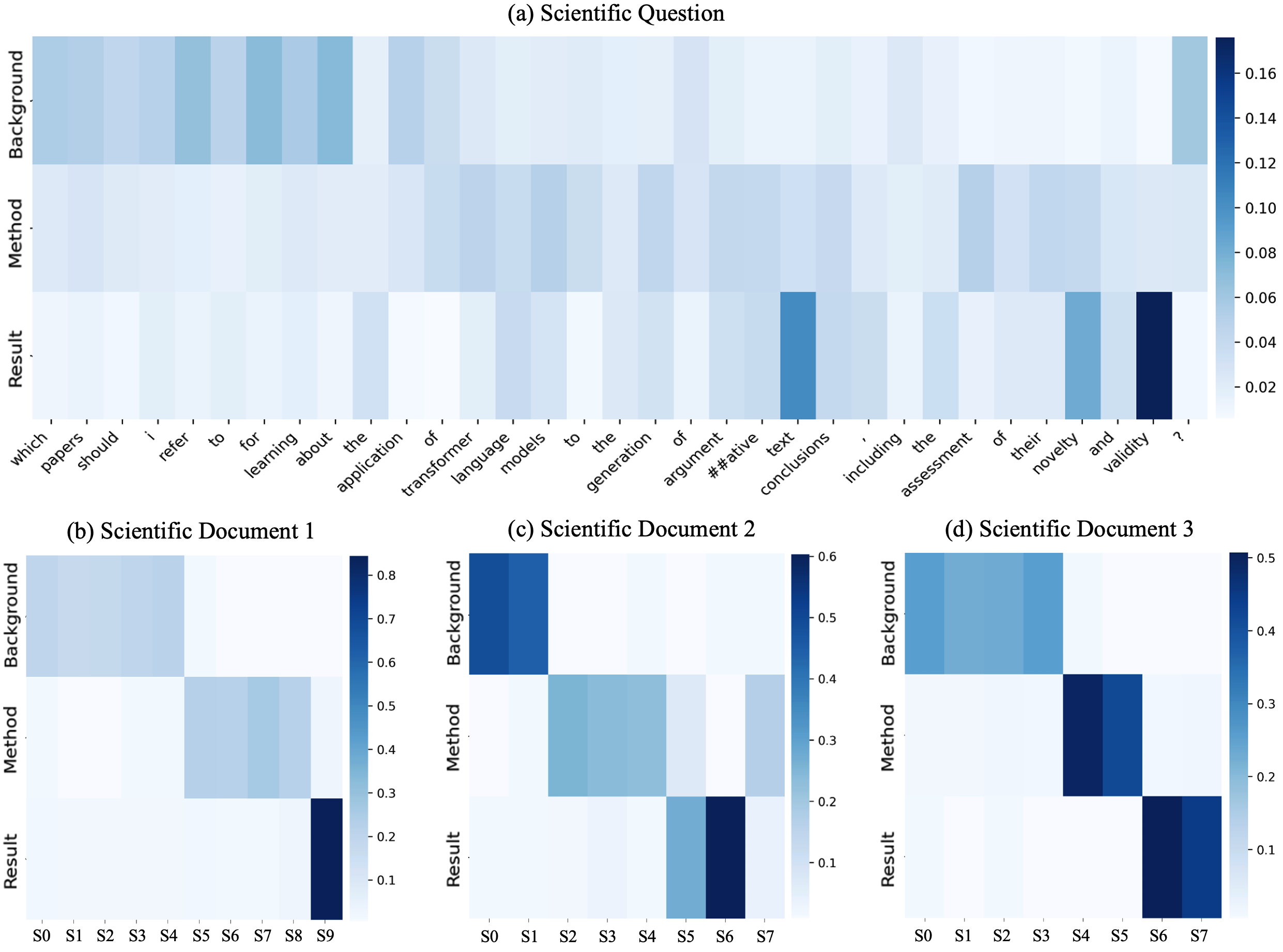}
  \caption{Facet-level attention visualizations for the scientific question and three retrieved documents in Table~\ref{tab:case-sample}. Each heatmap shows the attention weights from the three facet queries $\mathbf{Q}_{facet}$ to question tokens or document sentences during multi-granularity aggregation. Higher weights (darker colors) indicate greater contribution of a sentence or token to the final facet embeddings.}
  \label{fig_vis}  
\end{figure*}

{We visualize the facet-level attention distributions for a randomly selected sample from LitSearch to illustrate how facet embeddings capture different semantic aspects and yield interpretable representations.}
Table~\ref{tab:case-sample} provides detailed information for the question and retrieved documents, and Figure~\ref{fig_vis} shows the corresponding facet-level attention heatmaps.

\begin{table*}[t!]
  \centering
  \scriptsize
  \caption{An example consisting of one scientific question (top row) and three retrieved scientific documents (bottom row) from LitSearch. All abstract sentences are indexed (S0, S1, ...) for reference in the attention visualizations in Figure~\ref{fig_vis}.}
  \begin{tabular}{p{0.96\linewidth}}
  \toprule
  \textbf{Scientific Question} \\
  \midrule
  Which papers should I refer to for learning about the application of transformer language models to the generation of argumentative text conclusions, including the assessment of their novelty and validity? \\
  \bottomrule
  \end{tabular} 
  \begin{tabular}{p{0.3\linewidth} | p{0.3\linewidth} | p{0.3\linewidth}}
  \textbf{Scientific Document 1} & \textbf{Scientific Document 2} & \textbf{Scientific Document 3} \\
  \midrule
  \parbox[t]{\linewidth}{
  \textbf{Title:} Generating Informative Conclusions for Argumentative Texts\\
  \textbf{Abstract:} The purpose of an argumentative text is to support a certain conclusion. (S0)\\
  Yet, they are often omitted, expecting readers to infer them rather. (S1)\\
  While appropriate when reading an individual text, this rhetorical device limits accessibility when browsing many texts (e.g., on a search engine or on social media). (S2)\\
  In these scenarios, an explicit conclusion makes for a good candidate summary of an argumentative text. (S3)\\
  This is especially true if the conclusion is informative, emphasizing specific concepts from the text. (S4)\\
  With this paper we introduce the task of generating informative conclusions: First, Webis-ConcluGen-21 is compiled, a large-scale corpus of 136,996 samples of argumentative texts and their conclusions. (S5)\\
  Second, two paradigms for conclusion generation are investigated; one extractive, the other abstractive in nature. (S6)\\
  The latter exploits argumentative knowledge that augment the data via control codes and finetuning the BART model on several subsets of the corpus. (S7)\\
  Third, insights are provided into the suitability of our corpus for the task, the differences between the two generation paradigms, the trade-off between informativeness and conciseness, and the impact of encoding argumentative knowledge. (S8)\\
  The corpus, code, and the trained models are publicly available. (S9)
  } &
  \parbox[t]{\linewidth}{
  \textbf{Title:} Explainable Unsupervised Argument Similarity Rating with Abstract Meaning Representation and Conclusion Generation\\
  \textbf{Abstract:} When assessing the similarity of arguments, researchers typically use approaches that do not provide interpretable evidence or justifications for their ratings. (S0)\\
  Hence, the features that determine argument similarity remain elusive. (S1)\\
  We address this issue by introducing novel argument similarity metrics that aim at high performance and explainability. (S2)\\
  We show that Abstract Meaning Representation (AMR) graphs can be useful for representing arguments, and that novel AMR graph metrics can offer explanations for argument similarity ratings. (S3)\\
  We start from the hypothesis that similar premises often lead to similar conclusions and extend an approach for AMR-based argument similarity rating by estimating, in addition, the similarity of conclusions that we automatically infer from the arguments used as premises. (S4)\\
  We show that AMR similarity metrics make argument similarity judgements more interpretable and may even support argument quality judgements. (S5)\\
  Our approach provides significant performance improvements over strong baselines in a fully unsupervised setting. (S6)\\
  Finally, we make first steps to address the problem of reference-less evaluation of argumentative conclusion generations. (S7)
  } &
  \parbox[t]{\linewidth}{
  \textbf{Title:} Assessing the Sufficiency of Arguments through Conclusion Generation\\
  \textbf{Abstract:} The premises of an argument give evidence or other reasons to support a conclusion. (S0)\\
  However, the amount of support required depends on the generality of a conclusion, the nature of the individual premises, and similar. (S1)\\
  An argument whose premises make its conclusion rationally worthy to be drawn is called sufficient in argument quality research. (S2)\\
  Previous work tackled sufficiency assessment as a standard text classification problem, not modeling the inherent relation of premises and conclusion. (S3)\\
  In this paper, we hypothesize that the conclusion of a sufficient argument can be generated from its premises. (S4)\\
  To study this hypothesis, we explore the potential of assessing sufficiency based on the output of large-scale pre-trained language models. (S5)\\
  Our best model variant achieves an F1-score of 0.885, outperforming the previous state-of-the-art and being on par with human experts. (S6)\\
  While manual evaluation reveals the quality of the generated conclusions, their impact remains low ultimately. (S7)
  } \\
  \bottomrule
  \end{tabular}
  \label{tab:case-sample}
  \end{table*}

In Figure~\ref{fig_vis} (a), the scientific question is partitioned into semantically coherent zones by three facets. The \textit{Background} facet dominates the first third of the question, assigning substantial weight to the segment that establishes the informational scope, asking “which papers?” and specifying the topical focus on “the application”. This pattern indicates that UniFAR identifies these early tokens as defining the conceptual frame of the question. The \textit{Method} facet exhibits localized and moderate activations in the middle portion of the question, primarily around mentions of models or technical procedures such as “transformer language models” and “generation”. The \textit{Result} facet concentrates sharply on the evaluative tail of the question, with pronounced activation around “novelty” and “validity.” This suggests that the model interprets these terms as direct signals of result-level properties, consistent with their inherent role as evaluative criteria when formulating information needs. 

In Figure~\ref{fig_vis} (b)–(d), the three facet queries produce similarly coherent patterns across sentences of the retrieved documents. For each document, the \textit{Background} facet primarily attends to the opening sentences, where the abstracts introduce the research problem or conceptual motivation (Doc 1: S0–S4; Doc 2: S0–S1; Doc 3: S0–S3). The \textit{Method} facet shifts its focus to the middle portions of the abstracts, highlighting sentences that describe technical procedures or modeling choices (Doc 1: S5–S8; Doc 2: S2–S5,S7; Doc 3: S4–S5). The \textit{Result} facet consistently peaks near the end of each document, aligning with sentences that report empirical findings or performance outcomes (Doc 2: S5-S6; Doc 3: S6–S7).

These attention patterns show that UniFAR effectively disentangles background, method, and result information in both the question and the retrieved documents, yielding facet embeddings that correspond to distinct components of scientific discourse. The facet-specific signals track introductory context, procedural detail, and empirical findings aligned with the common structure of scientific writing. This alignment indicates that UniFAR’s multi-granularity aggregation does not diffuse attention uniformly but selectively focuses on segments that carry facet-relevant evidence. By grounding attention with facet anchors, UniFAR produces interpretable and stable facet embeddings that reflect semantic structure and question intent in realistic scientific document retrieval scenarios.

\subsection{{Efficiency Analysis (RQ6)}}

{We evaluate the efficiency of UniFAR on LitSearch across several
backbone scales. Table~\ref{tab:efficiency} reports wall-clock
latency for online retrieval, offline document encoding time,
peak GPU memory during online retrieval, parameter overhead, and
corpus vector storage.} {For offline encoding, we process all documents in the LitSearch
candidate pool with a batch size of 16 and divide the total elapsed time
by the number of documents to obtain the mean encoding time per
document. For online retrieval, we process one query at a time
and report the mean wall-clock latency for query encoding and
index search. To measure peak GPU memory, we run the SciBERT and Contriever
groups in FP32 and the Qwen3Emb groups in BF16, processing one
query at a time.}

\begin{table*}[t!]
  \centering
  \footnotesize
  \setlength{\tabcolsep}{3pt}
  \renewcommand{\arraystretch}{0.9}
  \caption{{Efficiency comparison on LitSearch across backbone scales.
  Online retrieval latency includes query encoding and index search. $|S|$ denotes the number of sentences in a document.}}
  \label{tab:efficiency}
  \begin{tabular}{@{}cccccccc@{}}
    \toprule
    & \multicolumn{3}{c}{{Runtime Efficiency}}
    & \multicolumn{2}{c}{Parameter Overhead}
    & \multicolumn{2}{c}{{Vector Storage}} \\
    \cmidrule(lr){2-4}\cmidrule(lr){5-6}\cmidrule(l){7-8}
    Model
    & {\shortstack{Offline Encoding\\Time (ms/doc)}}
    & {\shortstack{Online Retrieval\\Latency (ms/query)}}
    & {\shortstack{Peak GPU\\Memory (GB)}}
    & \shortstack{Extra\\Params (M)}
    & \shortstack{Relative\\Increase}
    & \#Vec / Doc
    & \shortstack{Storage\\(MB)} \\
    \midrule
    SciBERT{($\sim$110M)}
    & 6.28 & 8.36 & {0.84} & -- & -- & 1 & 168.92 \\
    {\scriptsize TS}Aspire
    & 19.95 & 128.95 & {0.82} & 0 & 0\% & $|S|$ & 999.25 \\
    FLeW
    & 18.70 & 29.85 & {1.79} & 220 & 200\% & 3 & 506.75 \\
    UniFAR{\tiny -SciBERT}
    & 7.04 & 16.89 & {0.86} & 6.50 & 5.91\% & 3 & 506.75 \\
    \midrule
    Contriever{($\sim$110M)}
    & 6.69 & 12.93 & {0.84} & -- & -- & 1 & 168.92 \\
    UniFAR{\tiny -Contriever}
    & 7.15 & 18.22 & {0.86} & 6.50 & {5.97\%} & 3 & 506.75 \\
    \midrule
    Qwen3Emb-0.6B
    & 11.93 & 44.78 & {1.57} & -- & -- & 1 & 225.22 \\
    UniFAR{\tiny -0.6B}
    & 12.96 & 54.69 & {1.59} & 11.55 & 1.93\% & 3 & 675.66 \\
    \midrule
    Qwen3Emb-4B
    & 49.57 & 94.33 & {8.50} & -- & -- & 1 & 563.06 \\
    UniFAR{\tiny -4B}
    & 51.12 & 106.21 & {8.66} & 72.13 & 1.80\% & 3 & 1689.18 \\
    \midrule
    Qwen3Emb-8B
    & 82.54 & 124.18 & {15.70} & -- & -- & 1 & 900.89 \\
    UniFAR{\tiny -8B}
    & 84.47 & 139.07 & {16.07} & 184.62 & 2.31\% & 3 & 2702.67 \\
    \bottomrule
  \end{tabular}
\end{table*}

\subsubsection{{Latency and Inference Memory.}} {UniFAR adds 0.46--1.93\,ms per document to offline encoding across
the tested backbones. The additional computation comes from the
adaptive aggregation used to construct facet vectors, while backbone
encoding remains the main cost. The increase is more noticeable
during online retrieval (5.29--14.89\,ms per query), where the system
also matches multiple facet vectors. 
UniFAR{\scriptsize -SciBERT} produces three facet vectors with a shared
encoder and achieves lower online retrieval latency than
{\scriptsize TS}Aspire (sentence-level matching) and FLeW
(separate facet encoders).} {Peak GPU memory increases by at most 0.37\,GB over the corresponding
backbones, consistent with sharing a single encoder and
running only one attention branch in the facet aggregation
module.}

\subsubsection{Parameter Overhead.}
{The parameter count of UniFAR's facet aggregation module grows with
the hidden size of the backbone. Although the absolute number of extra parameters increases with model scale, the relative increase remains low across all backbones, especially for larger models (1.80--2.31\% for the Qwen3Emb variants).} Compared with FLeW, which employs separate facet-specific encoders for different semantic facets, resulting in a substantial parameter increase (200\%), UniFAR adopts a unified representation design that enables facet-aware modeling with minimal parameter expansion.
This indicates that UniFAR does not rely on substantial parameter expansion to achieve better retrieval performance, but instead benefits from a lightweight and unified facet aggregation design.

\subsubsection{{Vector Storage.}}
{UniFAR adopts a facet-level multi-vector representation, storing a fixed number of embeddings per document, which places it between document-level and sentence-level methods. Compared to single-vector approaches, UniFAR{\scriptsize -SciBERT} requires more vector
storage (506.75 vs.\ 168.92\,MB for SciBERT), while using less storage
than a sentence-level method (506.75 vs.\ 999.25\,MB for
{\scriptsize TS}Aspire), which stores a variable number of vectors
per document. This design supports fine-grained facet matching
while keeping storage requirements controlled and predictable as the corpus grows.}

\section{Conclusion and Future Work}

In this paper, we present UniFAR, a unified facet-aware retrieval framework for scientific documents that jointly supports q-doc and doc-doc retrieval within a shared representation space. 
We identify systematic misalignments between these two paradigms in terms of input granularity, semantic complexity, and supervision signals, and address them through a unified modeling and training framework that integrates adaptive multi-granularity aggregation, facet-level modeling with learnable anchors, and facet-aware joint training. 
{Extensive experiments on multiple benchmark datasets across different scientific fields demonstrate that UniFAR consistently outperforms baselines under both q-doc and doc-doc retrieval settings as well as across different backbone models.} {Further analyses address research questions regarding facet matching strategies, ablation studies, attention visualization, and efficiency analysis, providing insights into the effectiveness of each component and the interpretability of facet-level representations.}

UniFAR establishes a unified framework for advancing scientific knowledge retrieval in the era of LLMs and RAG-based systems. 
Future work can explore leveraging facet-level representations for adaptive retrieval augmentation, controllable evidence selection, and fine-grained knowledge grounding in scientific generation tasks. 
{In addition, 
extending UniFAR to dynamically adapt facet structures according to query complexity, domain characteristics, and task requirements may further improve its flexibility and applicability in complex retrieval and generation scenarios.}

  
\bibliographystyle{ACM-Reference-Format}
\bibliography{sample-base}

@inproceedings{DPR,
    title = "Dense Passage Retrieval for Open-Domain Question Answering",
    author = "Karpukhin, Vladimir  and
      Oguz, Barlas  and
      Min, Sewon  and
      Lewis, Patrick  and
      Wu, Ledell  and
      Edunov, Sergey  and
      Chen, Danqi  and
      Yih, Wen-tau",
    editor = "Webber, Bonnie  and
      Cohn, Trevor  and
      He, Yulan  and
      Liu, Yang",
    booktitle = "Proceedings of the 2020 Conference on Empirical Methods in Natural Language Processing (EMNLP)",
    month = nov,
    year = "2020",
    address = "Online",
    publisher = "Association for Computational Linguistics",
    url = "https://aclanthology.org/2020.emnlp-main.550/",
    doi = "10.18653/v1/2020.emnlp-main.550",
    pages = "6769--6781"
}

@inproceedings{SimCSE,
    title = "{S}im{CSE}: Simple Contrastive Learning of Sentence Embeddings",
    author = "Gao, Tianyu  and
      Yao, Xingcheng  and
      Chen, Danqi",
    editor = "Moens, Marie-Francine  and
      Huang, Xuanjing  and
      Specia, Lucia  and
      Yih, Scott Wen-tau",
    booktitle = "Proceedings of the 2021 Conference on Empirical Methods in Natural Language Processing",
    month = nov,
    year = "2021",
    address = "Online and Punta Cana, Dominican Republic",
    publisher = "Association for Computational Linguistics",
    url = "https://aclanthology.org/2021.emnlp-main.552/",
    doi = "10.18653/v1/2021.emnlp-main.552",
    pages = "6894--6910"
}

@article{Contriever,
    title={Unsupervised Dense Information Retrieval with Contrastive Learning},
    author={Gautier Izacard and Mathilde Caron and Lucas Hosseini and Sebastian Riedel and Piotr Bojanowski and Armand Joulin and Edouard Grave},
    journal={Transactions on Machine Learning Research},
    issn={2835-8856},
    year={2022},
    url={https://openreview.net/forum?id=jKN1pXi7b0},
    note={}
}

@misc{E5,
    title={Text Embeddings by Weakly-Supervised Contrastive Pre-training}, 
    author={Liang Wang and Nan Yang and Xiaolong Huang and Binxing Jiao and Linjun Yang and Daxin Jiang and Rangan Majumder and Furu Wei},
    year={2024},
    eprint={2212.03533},
    archivePrefix={arXiv},
    primaryClass={cs.CL},
    url={https://arxiv.org/abs/2212.03533}, 
}

@inproceedings{specter,
    title = "{SPECTER}: Document-level Representation Learning using Citation-informed Transformers",
    author = "Cohan, Arman  and
      Feldman, Sergey  and
      Beltagy, Iz  and
      Downey, Doug  and
      Weld, Daniel",
    editor = "Jurafsky, Dan  and
      Chai, Joyce  and
      Schluter, Natalie  and
      Tetreault, Joel",
    booktitle = "Proceedings of the 58th Annual Meeting of the Association for Computational Linguistics",
    month = jul,
    year = "2020",
    address = "Online",
    publisher = "Association for Computational Linguistics",
    url = "https://aclanthology.org/2020.acl-main.207/",
    doi = "10.18653/v1/2020.acl-main.207",
    pages = "2270--2282"
}

@inproceedings{SciNCL,
    title = "Neighborhood Contrastive Learning for Scientific Document Representations with Citation Embeddings",
    author = "Ostendorff, Malte  and
      Rethmeier, Nils  and
      Augenstein, Isabelle  and
      Gipp, Bela  and
      Rehm, Georg",
    editor = "Goldberg, Yoav  and
      Kozareva, Zornitsa  and
      Zhang, Yue",
    booktitle = "Proceedings of the 2022 Conference on Empirical Methods in Natural Language Processing",
    month = dec,
    year = "2022",
    address = "Abu Dhabi, United Arab Emirates",
    publisher = "Association for Computational Linguistics",
    url = "https://aclanthology.org/2022.emnlp-main.802/",
    doi = "10.18653/v1/2022.emnlp-main.802",
    pages = "11670--11688"
}

@inproceedings{FLeW,
  title     = {{FLeW}: Facet-Level and Adaptive Weighted Representation Learning of Scientific Documents},
  author    = {Zheng Dou and Deqing Wang and Fuzhen Zhuang and Jian Ren and Yanlin Hu},
  booktitle = {Database Systems for Advanced Applications -- 30th International Conference, {DASFAA} 2025, Singapore, May 26--29, 2025, Proceedings, Part I},
  series    = {Lecture Notes in Computer Science},
  volume    = {15986},
  pages     = {317--327},
  year      = {2026},
  publisher = {Springer},
  address   = {Singapore},
  doi       = {10.1007/978-981-95-3827-0_22},
  url       = {https://doi.org/10.1007/978-981-95-3827-0_22}
}

@inproceedings{FaBle,
    title = "Multi-Facet Blending for Faceted Query-by-Example Retrieval",
    author = "Do, Heejin  and
      Ryu, Sangwon  and
      Kim, Jonghwi  and
      Lee, Gary",
    editor = "Che, Wanxiang  and
      Nabende, Joyce  and
      Shutova, Ekaterina  and
      Pilehvar, Mohammad Taher",
    booktitle = "Proceedings of the 63rd Annual Meeting of the Association for Computational Linguistics (Volume 1: Long Papers)",
    month = jul,
    year = "2025",
    address = "Vienna, Austria",
    publisher = "Association for Computational Linguistics",
    url = "https://aclanthology.org/2025.acl-long.1388/",
    doi = "10.18653/v1/2025.acl-long.1388",
    pages = "28577--28590",
    ISBN = "979-8-89176-251-0"
}

@inproceedings{specter2,
    title = "{S}ci{R}ep{E}val: A Multi-Format Benchmark for Scientific Document Representations",
    author = "Singh, Amanpreet  and
      D{'}Arcy, Mike  and
      Cohan, Arman  and
      Downey, Doug  and
      Feldman, Sergey",
    editor = "Bouamor, Houda  and
      Pino, Juan  and
      Bali, Kalika",
    booktitle = "Proceedings of the 2023 Conference on Empirical Methods in Natural Language Processing",
    month = dec,
    year = "2023",
    address = "Singapore",
    publisher = "Association for Computational Linguistics",
    url = "https://aclanthology.org/2023.emnlp-main.338/",
    doi = "10.18653/v1/2023.emnlp-main.338",
    pages = "5548--5566"
}

@inproceedings{aspire,
    title = "Multi-Vector Models with Textual Guidance for Fine-Grained Scientific Document Similarity",
    author = "Mysore, Sheshera  and
      Cohan, Arman  and
      Hope, Tom",
    editor = "Carpuat, Marine  and
      de Marneffe, Marie-Catherine  and
      Meza Ruiz, Ivan Vladimir",
    booktitle = "Proceedings of the 2022 Conference of the North American Chapter of the Association for Computational Linguistics: Human Language Technologies",
    month = jul,
    year = "2022",
    address = "Seattle, United States",
    publisher = "Association for Computational Linguistics",
    url = "https://aclanthology.org/2022.naacl-main.331/",
    doi = "10.18653/v1/2022.naacl-main.331",
    pages = "4453--4470"
}

@article{OpenScholar,
    title={Synthesizing scientific literature with retrieval-augmented language models}, 
    author={Akari Asai and Jacqueline He and Rulin Shao and Weijia Shi and Amanpreet Singh and Joseph Chee Chang and Kyle Lo and Luca Soldaini and Sergey Feldman and Mike D'arcy and David Wadden and Matt Latzke and Jenna Sparks and Jena Hwang and Varsha Kishore and Minyang Tian and Pan Ji and Shengyan Liu and Hao Tong and Bohao Wu and Yanyu Xiong and Luke Zettlemoyer and Graham Neubig and Dan Weld and Doug Downey and Wen-tau Yih and Pang Wei Koh and Hannaneh Hajishirzi},
    journal={Nature},
    volume={650},
    pages={857–-863},
    year={2026},
    doi={10.1038/s41586-025-10072-4},
    url={https://www.nature.com/articles/s41586-025-10072-4#citeas}
}

@inproceedings{OpenResearcher,
    title = "{O}pen{R}esearcher: Unleashing {AI} for Accelerated Scientific Research",
    author = "Zheng, Yuxiang  and
      Sun, Shichao  and
      Qiu, Lin  and
      Ru, Dongyu  and
      Jiayang, Cheng  and
      Li, Xuefeng  and
      Lin, Jifan  and
      Wang, Binjie  and
      Luo, Yun  and
      Pan, Renjie  and
      Xu, Yang  and
      Min, Qingkai  and
      Zhang, Zizhao  and
      Wang, Yiwen  and
      Li, Wenjie  and
      Liu, Pengfei",
    editor = "Hernandez Farias, Delia Irazu  and
      Hope, Tom  and
      Li, Manling",
    booktitle = "Proceedings of the 2024 Conference on Empirical Methods in Natural Language Processing: System Demonstrations",
    month = nov,
    year = "2024",
    address = "Miami, Florida, USA",
    publisher = "Association for Computational Linguistics",
    url = "https://aclanthology.org/2024.emnlp-demo.22/",
    doi = "10.18653/v1/2024.emnlp-demo.22",
    pages = "209--218"
}

@article{RAG,
    title={Retrieval-augmented generation for knowledge-intensive nlp tasks},
    author={Lewis, Patrick and Perez, Ethan and Piktus, Aleksandra and Petroni, Fabio and Karpukhin, Vladimir and Goyal, Naman and K{\"u}ttler, Heinrich and Lewis, Mike and Yih, Wen-tau and Rockt{\"a}schel, Tim and others},
    journal={Advances in neural information processing systems},
    volume={33},
    pages={9459--9474},
    year={2020},
    url={https://proceedings.neurips.cc/paper/2020/file/6b493230205f780e1bc26945df7481e5-Paper.pdf}
}

@inproceedings{RAG-LLM,
    author = {Fan, Wenqi and Ding, Yujuan and Ning, Liangbo and Wang, Shijie and Li, Hengyun and Yin, Dawei and Chua, Tat-Seng and Li, Qing},
    title = {A Survey on RAG Meeting LLMs: Towards Retrieval-Augmented Large Language Models},
    year = {2024},
    isbn = {9798400704901},
    publisher = {Association for Computing Machinery},
    address = {New York, NY, USA},
    url = {https://doi.org/10.1145/3637528.3671470},
    doi = {10.1145/3637528.3671470},
    booktitle = {Proceedings of the 30th ACM SIGKDD Conference on Knowledge Discovery and Data Mining},
    pages = {6491–6501},
    numpages = {11},
    location = {Barcelona, Spain},
    series = {KDD '24}
}

@misc{InfoNCE,
    title={Representation Learning with Contrastive Predictive Coding}, 
    author={{Oord}, Aaron van den and Yazhe Li and Oriol Vinyals},
    year={2019},
    eprint={1807.03748},
    archivePrefix={arXiv},
    primaryClass={cs.LG},
    url={https://arxiv.org/abs/1807.03748}, 
}

@INPROCEEDINGS{margin,
    author={Schroff, Florian and Kalenichenko, Dmitry and Philbin, James},
    booktitle={2015 IEEE Conference on Computer Vision and Pattern Recognition (CVPR)}, 
    title={FaceNet: A unified embedding for face recognition and clustering}, 
    year={2015},
    volume={},
    number={},
    pages={815-823},
    doi={10.1109/CVPR.2015.7298682}
}

@inproceedings{scibert,
    title = "{S}ci{BERT}: A Pretrained Language Model for Scientific Text",
    author = "Beltagy, Iz  and
      Lo, Kyle  and
      Cohan, Arman",
    editor = "Inui, Kentaro  and
      Jiang, Jing  and
      Ng, Vincent  and
      Wan, Xiaojun",
    booktitle = "Proceedings of the 2019 Conference on Empirical Methods in Natural Language Processing and the 9th International Joint Conference on Natural Language Processing (EMNLP-IJCNLP)",
    month = nov,
    year = "2019",
    address = "Hong Kong, China",
    publisher = "Association for Computational Linguistics",
    url = "https://aclanthology.org/D19-1371/",
    doi = "10.18653/v1/D19-1371",
    pages = "3615--3620"
}

@misc{qwen,
    title={Qwen3 Embedding: Advancing Text Embedding and Reranking Through Foundation Models}, 
    author={Yanzhao Zhang and Mingxin Li and Dingkun Long and Xin Zhang and Huan Lin and Baosong Yang and Pengjun Xie and An Yang and Dayiheng Liu and Junyang Lin and Fei Huang and Jingren Zhou},
    year={2025},
    eprint={2506.05176},
    archivePrefix={arXiv},
    primaryClass={cs.CL},
    url={https://arxiv.org/abs/2506.05176}, 
}

@article{dorismae,
  title={Scientific document retrieval using multi-level aspect-based queries},
  author={Wang, Jianyou Andre and Wang, Kaicheng and Wang, Xiaoyue and Naidu, Prudhviraj and Bergen, Leon and Paturi, Ramamohan},
  journal={Advances in Neural Information Processing Systems},
  volume={36},
  pages={38404--38419},
  year={2023},
  url="https://proceedings.neurips.cc/paper_files/paper/2023/file/78f9c04bdcb06f1ada3902912d8b64ba-Paper-Datasets_and_Benchmarks.pdf"
}

@inproceedings{csfcube,
    title={{CSFC}ube - A Test Collection of Computer Science Research Articles for Faceted Query by Example},
    author={Sheshera Mysore and Tim O'Gorman and Andrew McCallum and Hamed Zamani},
    booktitle={Thirty-fifth Conference on Neural Information Processing Systems Datasets and Benchmarks Track},
    year={2021},
    url={https://datasets-benchmarks-proceedings.neurips.cc/paper_files/paper/2021/file/20f07591c6fcb220ffe637cda29bb3f6-Paper-round2.pdf}
}

@inproceedings{litsearch,
    title = "{L}it{S}earch: A Retrieval Benchmark for Scientific Literature Search",
    author = "Ajith, Anirudh  and
      Xia, Mengzhou  and
      Chevalier, Alexis  and
      Goyal, Tanya  and
      Chen, Danqi  and
      Gao, Tianyu",
    editor = "Al-Onaizan, Yaser  and
      Bansal, Mohit  and
      Chen, Yun-Nung",
    booktitle = "Proceedings of the 2024 Conference on Empirical Methods in Natural Language Processing",
    month = nov,
    year = "2024",
    address = "Miami, Florida, USA",
    publisher = "Association for Computational Linguistics",
    url = "https://aclanthology.org/2024.emnlp-main.840/",
    doi = "10.18653/v1/2024.emnlp-main.840",
    pages = "15068--15083"
}

@inproceedings{MTEB,
    title = "{MTEB}: Massive Text Embedding Benchmark",
    author = "Muennighoff, Niklas  and
      Tazi, Nouamane  and
      Magne, Loic  and
      Reimers, Nils",
    editor = "Vlachos, Andreas  and
      Augenstein, Isabelle",
    booktitle = "Proceedings of the 17th Conference of the European Chapter of the Association for Computational Linguistics",
    month = may,
    year = "2023",
    address = "Dubrovnik, Croatia",
    publisher = "Association for Computational Linguistics",
    url = "https://aclanthology.org/2023.eacl-main.148/",
    doi = "10.18653/v1/2023.eacl-main.148",
    pages = "2014--2037"
}

@inproceedings{SciRAG,
    title = "{S}ci{RAG}: Adaptive, Citation-Aware, and Outline-Guided Retrieval and Synthesis for Scientific Literature",
    author = "Ding, Hang  and
      Zhao, Yilun  and
      Hu, Tiansheng  and
      Wang, Zihang  and
      Patwardhan, Manasi  and
      Cohan, Arman",
    editor = "Demberg, Vera  and
      Inui, Kentaro  and
      Marquez, Llu{\'i}s",
    booktitle = "Proceedings of the 19th Conference of the {E}uropean Chapter of the {A}ssociation for {C}omputational {L}inguistics (Volume 1: Long Papers)",
    month = mar,
    year = "2026",
    address = "Rabat, Morocco",
    publisher = "Association for Computational Linguistics",
    url = "https://aclanthology.org/2026.eacl-long.303/",
    doi = "10.18653/v1/2026.eacl-long.303",
    pages = "6440--6460",
    ISBN = "979-8-89176-380-7"
}

@inproceedings{mixgr,
    title = "{M}ix{GR}: Enhancing Retriever Generalization for Scientific Domain through Complementary Granularity",
    author = "Cai, Fengyu  and
      Zhao, Xinran  and
      Chen, Tong  and
      Chen, Sihao  and
      Zhang, Hongming  and
      Gurevych, Iryna  and
      Koeppl, Heinz",
    editor = "Al-Onaizan, Yaser  and
      Bansal, Mohit  and
      Chen, Yun-Nung",
    booktitle = "Proceedings of the 2024 Conference on Empirical Methods in Natural Language Processing",
    month = nov,
    year = "2024",
    address = "Miami, Florida, USA",
    publisher = "Association for Computational Linguistics",
    url = "https://aclanthology.org/2024.emnlp-main.579/",
    doi = "10.18653/v1/2024.emnlp-main.579",
    pages = "10369--10391"
}

@inproceedings{CCQGen,
  title = {Improving Scientific Document Retrieval with Concept Coverage-based Query Set Generation},
  author = {Kang, SeongKu and Jin, Bowen and Kweon, Wonbin and Zhang, Yu and Lee, Dongha and Han, Jiawei and Yu, Hwanjo},
  year = {2025},
  isbn = {9798400713293},
  publisher = {Association for Computing Machinery},
  address = {New York, NY, USA},
  url = {https://doi.org/10.1145/3701551.3703544},
  doi = {10.1145/3701551.3703544},
  booktitle = {Proceedings of the Eighteenth ACM International Conference on Web Search and Data Mining},
  pages = {895–904},
  numpages = {10},
  location = {Hannover, Germany},
  series = {WSDM '25}
}

@misc{CCExpand,
  title={Improving Scientific Document Retrieval with Academic Concept Index}, 
  author={Jeyun Lee and Junhyoung Lee and Wonbin Kweon and Bowen Jin and Yu Zhang and Susik Yoon and Dongha Lee and Hwanjo Yu and Jiawei Han and Seongku Kang},
  year={2026},
  eprint={2601.00567},
  archivePrefix={arXiv},
  primaryClass={cs.IR},
  url={https://arxiv.org/abs/2601.00567}, 
}

@misc{ai4r,
  title={AI4Research: A Survey of Artificial Intelligence for Scientific Research}, 
  author={Qiguang Chen and Mingda Yang and Libo Qin and Jinhao Liu and Zheng Yan and Jiannan Guan and Dengyun Peng and Yiyan Ji and Hanjing Li and Mengkang Hu and Yimeng Zhang and Yihao Liang and Yuhang Zhou and Jiaqi Wang and Zhi Chen and Wanxiang Che},
  year={2025},
  eprint={2507.01903},
  archivePrefix={arXiv},
  primaryClass={cs.CL},
  url={https://arxiv.org/abs/2507.01903}, 
}

@inproceedings{bert,
    title = "{BERT}: Pre-training of Deep Bidirectional Transformers for Language Understanding",
    author = "Devlin, Jacob  and
      Chang, Ming-Wei  and
      Lee, Kenton  and
      Toutanova, Kristina",
    editor = "Burstein, Jill  and
      Doran, Christy  and
      Solorio, Thamar",
    booktitle = "Proceedings of the 2019 Conference of the North {A}merican Chapter of the Association for Computational Linguistics: Human Language Technologies, Volume 1 (Long and Short Papers)",
    month = jun,
    year = "2019",
    address = "Minneapolis, Minnesota",
    publisher = "Association for Computational Linguistics",
    url = "https://aclanthology.org/N19-1423/",
    doi = "10.18653/v1/N19-1423",
    pages = "4171--4186"
}

@article{mpnet,
  title={Mpnet: Masked and permuted pre-training for language understanding},
  author={Song, Kaitao and Tan, Xu and Qin, Tao and Lu, Jianfeng and Liu, Tie-Yan},
  journal={Advances in neural information processing systems},
  volume={33},
  pages={16857--16867},
  year={2020},
  url = "https://proceedings.neurips.cc/paper_files/paper/2020/file/c3a690be93aa602ee2dc0ccab5b7b67e-Paper.pdf"
}

@inproceedings{sbert,
  title = "Sentence-{BERT}: Sentence Embeddings using {S}iamese {BERT}-Networks",
  author = "Reimers, Nils  and
    Gurevych, Iryna",
  editor = "Inui, Kentaro  and
    Jiang, Jing  and
    Ng, Vincent  and
    Wan, Xiaojun",
  booktitle = "Proceedings of the 2019 Conference on Empirical Methods in Natural Language Processing and the 9th International Joint Conference on Natural Language Processing (EMNLP-IJCNLP)",
  month = nov,
  year = "2019",
  address = "Hong Kong, China",
  publisher = "Association for Computational Linguistics",
  url = "https://aclanthology.org/D19-1410/",
  doi = "10.18653/v1/D19-1410",
  pages = "3982--3992"
}

@inproceedings{declutr,
    title = "{D}e{CLUTR}: Deep Contrastive Learning for Unsupervised Textual Representations",
    author = "Giorgi, John  and
      Nitski, Osvald  and
      Wang, Bo  and
      Bader, Gary",
    editor = "Zong, Chengqing  and
      Xia, Fei  and
      Li, Wenjie  and
      Navigli, Roberto",
    booktitle = "Proceedings of the 59th Annual Meeting of the Association for Computational Linguistics and the 11th International Joint Conference on Natural Language Processing (Volume 1: Long Papers)",
    month = aug,
    year = "2021",
    address = "Online",
    publisher = "Association for Computational Linguistics",
    url = "https://aclanthology.org/2021.acl-long.72/",
    doi = "10.18653/v1/2021.acl-long.72",
    pages = "879--895"
}

@inproceedings{BEIR,
    title={BEIR: A Heterogenous Benchmark for Zero-shot Evaluation of Information Retrieval Models}, 
    author={Nandan Thakur and Nils Reimers and Andreas Rücklé and Abhishek Srivastava and Iryna Gurevych},
    booktitle={Thirty-fifth Conference on Neural Information Processing Systems Datasets and Benchmarks Track},
    year={2021},
    url={https://datasets-benchmarks-proceedings.neurips.cc/paper/2021/file/65b9eea6e1cc6bb9f0cd2a47751a186f-Paper-round2.pdf}
}

@article{biobert,
  title={BioBERT: a pre-trained biomedical language representation model for biomedical text mining},
  author={Lee, Jinhyuk and Yoon, Wonjin and Kim, Sungdong and Kim, Donghyeon and Kim, Sunkyu and So, Chan Ho and Kang, Jaewoo},
  journal={Bioinformatics},
  volume={36},
  number={4},
  pages={1234--1240},
  year={2020},
  publisher={Oxford University Press},
  url={https://academic.oup.com/bioinformatics/article/36/4/1234/5566506},
}

@inproceedings{MIR,
    title = "{MIR}: Methodology Inspiration Retrieval for Scientific Research Problems",
    author = "Garikaparthi, Aniketh  and
      Patwardhan, Manasi  and
      Kanade, Aditya Sanjiv  and
      Hassan, Aman  and
      Vig, Lovekesh  and
      Cohan, Arman",
    editor = "Che, Wanxiang  and
      Nabende, Joyce  and
      Shutova, Ekaterina  and
      Pilehvar, Mohammad Taher",
    booktitle = "Proceedings of the 63rd Annual Meeting of the Association for Computational Linguistics (Volume 1: Long Papers)",
    month = jul,
    year = "2025",
    address = "Vienna, Austria",
    publisher = "Association for Computational Linguistics",
    url = "https://aclanthology.org/2025.acl-long.1390/",
    doi = "10.18653/v1/2025.acl-long.1390",
    pages = "28614--28659",
    ISBN = "979-8-89176-251-0"
}

@inproceedings{beyondcitations,
    title = "Beyond Citations: Integrating Finding-Based Relations for Improved Biomedical Article Representations",
    author = "Liang, Yuan  and
      Poesio, Massimo  and
      Rezvani, Roonak",
    editor = "Demner-Fushman, Dina  and
      Ananiadou, Sophia  and
      Miwa, Makoto  and
      Tsujii, Junichi",
    booktitle = "Proceedings of the 24th Workshop on Biomedical Language Processing",
    month = aug,
    year = "2025",
    address = "Viena, Austria",
    publisher = "Association for Computational Linguistics",
    url = "https://aclanthology.org/2025.bionlp-1.25/",
    doi = "10.18653/v1/2025.bionlp-1.25",
    pages = "297--306",
    ISBN = "979-8-89176-275-6"
}

@misc{paperqa,
  title={PaperQA: Retrieval-Augmented Generative Agent for Scientific Research}, 
  author={Jakub Lála and Odhran O'Donoghue and Aleksandar Shtedritski and Sam Cox and Samuel G. Rodriques and Andrew D. White},
  year={2023},
  eprint={2312.07559},
  archivePrefix={arXiv},
  primaryClass={cs.CL},
  url={https://arxiv.org/abs/2312.07559}, 
}

@inproceedings{qasper,
    title = "A Dataset of Information-Seeking Questions and Answers Anchored in Research Papers",
    author = "Dasigi, Pradeep  and
      Lo, Kyle  and
      Beltagy, Iz  and
      Cohan, Arman  and
      Smith, Noah A.  and
      Gardner, Matt",
    editor = "Toutanova, Kristina  and
      Rumshisky, Anna  and
      Zettlemoyer, Luke  and
      Hakkani-Tur, Dilek  and
      Beltagy, Iz  and
      Bethard, Steven  and
      Cotterell, Ryan  and
      Chakraborty, Tanmoy  and
      Zhou, Yichao",
    booktitle = "Proceedings of the 2021 Conference of the North American Chapter of the Association for Computational Linguistics: Human Language Technologies",
    month = jun,
    year = "2021",
    address = "Online",
    publisher = "Association for Computational Linguistics",
    url = "https://aclanthology.org/2021.naacl-main.365/",
    doi = "10.18653/v1/2021.naacl-main.365",
    pages = "4599--4610"
}

@inproceedings{qasa, 
  title = {QASA: advanced question answering on scientific articles},
  author = {Lee, Yoonjoo and Lee, Kyungjae and Park, Sunghyun and Hwang, Dasol and Kim, Jaehyeon and Lee, Hong-in and Lee, Moontae},
  year = {2023},
  publisher = {JMLR.org},
  booktitle = {Proceedings of the 40th International Conference on Machine Learning},
  articleno = {787},
  numpages = {17},
  location = {Honolulu, Hawaii, USA},
  series = {ICML'23},
  url = {https://dl.acm.org/doi/10.5555/3618408.3619195}
}

@misc{llm4research,
  title={LLM4SR: A Survey on Large Language Models for Scientific Research}, 
  author={Ziming Luo and Zonglin Yang and Zexin Xu and Wei Yang and Xinya Du},
  year={2025},
  eprint={2501.04306},
  archivePrefix={arXiv},
  primaryClass={cs.CL},
  url={https://arxiv.org/abs/2501.04306}, 
}

@inproceedings{limitations,
  title={On the Theoretical Limitations of Embedding-Based Retrieval}, 
  author={Orion Weller and Michael Boratko and Iftekhar Naim and Jinhyuk Lee},
  booktitle={International Conference on Learning Representations},
  year={2026},
  url={https://openreview.net/pdf?id=k9CzIvzfaA}
}

@inproceedings{scimult,
  title = "Pre-training Multi-task Contrastive Learning Models for Scientific Literature Understanding",
  author = "Zhang, Yu  and
    Cheng, Hao  and
    Shen, Zhihong  and
    Liu, Xiaodong  and
    Wang, Ye-Yi  and
    Gao, Jianfeng",
  editor = "Bouamor, Houda  and
    Pino, Juan  and
    Bali, Kalika",
  booktitle = "Findings of the Association for Computational Linguistics: EMNLP 2023",
  month = dec,
  year = "2023",
  address = "Singapore",
  publisher = "Association for Computational Linguistics",
  url = "https://aclanthology.org/2023.findings-emnlp.820/",
  doi = "10.18653/v1/2023.findings-emnlp.820",
  pages = "12259--12275"
}

@article{relish,
  author  = {Brown, Peter and Kulkarni, Ameya Sadguru and Refai, Osama and Zhou, Yaoqi},
  title   = {Large expert-curated database for benchmarking document
             similarity detection in biomedical literature search},
  journal = {Database: The Journal of Biological Databases and Curation},
  year    = {2019},
  volume  = {2019},
  articleno = {baz085},
  doi     = {10.1093/database/baz085}
}

@article{treccovid,
  author    = {Voorhees, Ellen and Alam, Tasmeer and Bedrick, Steven
               and Demner-Fushman, Dina and Hersh, William R. and Lo, Kyle
               and Roberts, Kirk and Soboroff, Ian and Wang, Lucy Lu},
  title     = {{TREC-COVID}: Constructing a Pandemic Information
               Retrieval Test Collection},
  journal   = {ACM SIGIR Forum},
  year      = {2021},
  month     = feb,
  volume    = {54},
  number    = {1},
  articleno = {1},
  numpages  = {12},
  doi       = {10.1145/3451964.3451965}
}

@inproceedings{BGE,
  author    = {Xiao, Shitao and Liu, Zheng and Zhang, Peitian and
               Muennighoff, Niklas and Lian, Defu and Nie, Jian-Yun},
  title     = {{C-Pack}: Packed Resources For General {Chinese} Embeddings},
  booktitle = {Proceedings of the 47th International ACM SIGIR Conference
               on Research and Development in Information Retrieval},
  year      = {2024},
  pages     = {641--649},
  publisher = {Association for Computing Machinery},
  doi       = {10.1145/3626772.3657878}
}

@inproceedings{gritlm,
  title     = {Generative Representational Instruction Tuning},
  author    = {Muennighoff, Niklas and Su, Hongjin and Wang, Liang
               and Yang, Nan and Wei, Furu and Yu, Tao
               and Singh, Amanpreet and Kiela, Douwe},
  booktitle = {The Thirteenth International Conference on Learning Representations},
  year      = {2025},
  url       = {https://openreview.net/forum?id=BC4lIvfSzv}
}

\appendix
\section{Evaluation Metrics}
\label{app:metrics}
We provide the formal definitions of the evaluation metrics used in our experiments. Given a query $q$, let $\mathcal{R}_q$ denote the set of relevant documents and $\pi_k(q)$ denote the top-$k$ ranked results.

\paragraph{Recall@$k$}
Recall@$k$ measures the fraction of relevant documents retrieved in the top-$k$ results:
\begin{equation}
\text{Recall@}k = \frac{|\mathcal{R}_q \cap \pi_k(q)|}{|\mathcal{R}_q|}.
\end{equation}

\paragraph{Mean Average Precision (MAP)}
MAP averages the average precision (AP) across queries:
\begin{equation}
\text{MAP}
= \frac{1}{|Q|} \sum_{q \in Q} \text{AP}(q)
= \frac{1}{|Q|} \sum_{q \in Q}
  \frac{1}{|\mathcal{R}_q|} \sum_{i=1}^{N}
  \text{Prec@}i(q) \cdot \mathbb{I}(d_i(q) \in \mathcal{R}_q).
\end{equation}

\paragraph{nDCG}
The normalized Discounted Cumulative Gain at cutoff $k$ is defined as:
\begin{equation}
\text{nDCG@}k = \frac{1}{Z_k} \sum_{i=1}^{k} \frac{2^{rel_i} - 1}{\log_2(i+1)},
\end{equation}
where $rel_i$ denotes the graded relevance at rank $i$, and $Z_k$ is a normalization factor. 
For CSFCube, nDCG$_{\%20}$ is computed over the top 20\% of the ranked list.

\paragraph{R-Precision}
R-Precision is defined as the precision at rank $|\mathcal{R}_q|$:
\begin{equation}
\text{R-Precision} = \text{Prec@}|\mathcal{R}_q|.
\end{equation}

\section{Prompt Templates Used for Training Data Construction}
\label{sec:prompts}
  
Table~\ref{tab:question} and Table~\ref{tab:document} provide the LLM prompt templates used in our training data construction pipeline (Section~\ref{data construction}) for question generation and {sentence labeling} {by Qwen3-Max-2025-09-23.} 

\begin{table}[h!]
  \caption{Prompt template used for generating facet-specific scientific questions from the query {paper} and its facet-positive {papers}.}
  \centering
  \footnotesize
  \renewcommand{\arraystretch}{1.0}
  \begin{tabular}{p{0.99\linewidth}}
  \toprule
  \textbf{System Message} \\
  \midrule
  You are an expert in scientific literature retrieval, specializing in generating professional academic search queries.
  \\[2mm]
  \toprule
  \textbf{User Prompt} \\
  \midrule
  Generate an English query for the \{\textit{facet\_type}\} facet based on the following sentences from the seed paper and positive papers.

  \{\textit{facet\_type}\} sentences from seed paper:  
  \{\texttt{query\_text}\}

  \{\textit{facet\_type}\} sentences from positive papers:  
  \{\texttt{pos\_text}\}

  Requirements:

  1.~Generate a natural, fluent, concise, and professional query that highlights \{\textit{requirement}\}.  

  2.~Ground the query in the provided sentences and avoid introducing unsupported concepts or assumptions. 

  3.~Output only the query itself, without any explanation or additional content.

  Generated \{\textit{facet\_type}\} query: \\
  \midrule
  \end{tabular}
  \begin{tabular}{lp{0.82\linewidth}}
  \{\textit{facet\_type}\} & \{\textit{requirement}\} \\
  \midrule
  background & research questions, research motivation, and limitations of existing studies \\
  method     & technical methods, experimental design, and innovations \\
  result     & key findings, data support, and conclusions \\
  \bottomrule
  \end{tabular}
  \label{tab:question}
\end{table}

\begin{table}[!h]
  \caption{Prompt template used for facet-level sentence labeling in scientific abstracts.}
  \centering
  \renewcommand{\arraystretch}{1.0}
  \footnotesize
  \begin{tabular}{p{0.99\linewidth}}
  \toprule
  \textbf{System Message} \\
  \midrule
  You are an expert in classifying sentences from scientific paper abstracts into rhetorical roles.
  \\[2mm]
  \toprule
  \textbf{User Prompt} \\
  \midrule
  The following sentences come from the same scientific abstract:\{\texttt{sentences\_list}\}

  Task Description:

  1.~Based on the content and contextual meaning, classify each sentence into one of the following categories: \{background, method, result\}.  

  2.~Return the output in a structured JSON format, ensuring each sentence is paired with the correct category.  

  3.~The expected output format is:

  [\{ "sentence": "Sentence1", "category": "background" \},

  \{ "sentence": "Sentence2", "category": "method" \},

  \{ "sentence": "Sentence3", "category": "result" \}]

  Please output only the structured JSON result without any explanation.
  \\
  \bottomrule
  \end{tabular}
  \label{tab:document}
\end{table}

\section{{An example from Facet-Aware Training Units (FTUs)}}
\label{sec:ftu-example}
{Table~\ref{tab:ftu-example} illustrates the JSON structure of an example from our constructed FTUs. The \texttt{query\_id} identifies $d_q$, while \texttt{positive\_ids},\texttt{negative\_ids}, and \texttt{questions} specify $\mathcal{D}_{\mathrm{pos}}^{(f)}$, $\mathcal{D}_{\mathrm{neg}}^{(f)}$, and $q^{(f)}$ for each facet $f$ (Eq.~\ref{eq:ftu}). Paper identifiers link the query and candidate sets to entries in \texttt{metadata}, which store each paper's title, abstract sentences, and facet labels. Sentences in \texttt{abstract\_sentences} retain their original order in the abstract. The \texttt{abstract\_sentences} and \texttt{facet\_labels} arrays are aligned by position.}

\begin{table*}[h!]
\centering
\caption{{JSON structure of an example FTU. Facet-specific paper sets and questions are stored separately from paper metadata and linked by paper identifiers. Metadata of other papers are omitted for brevity.}
}
\label{tab:ftu-example}
\scriptsize
\renewcommand{\arraystretch}{1.1}
\begin{tabular}{p{0.96\linewidth}}
\toprule
{\{} \\

{\hspace*{1em}"query\_id": "2621465",} \\

{\hspace*{1em}"positive\_ids": \{} \\
{\hspace*{2em}"background": ["368524", "65083087", "1345",
"15664890", "3146611", "218977368", "11599080",
"18796136", "6493988", "5060957"],} \\
{\hspace*{2em}"method": ["6946103", "69833386", "10168768",
"15218192", "231846372", "198905491", "231740694",
"193718462", "1916754", "52338594"],} \\
{\hspace*{2em}"result": ["23100864", "14223", "195064593",
"11403097", "6455915", "15307333", "219938042",
"15407650", "14747729", "149368"]} \\
{\hspace*{1em}\},} \\

{\hspace*{1em}"negative\_ids": \{} \\
{\hspace*{2em}"background": ["11588315", "7860810", "1549343",
"10913456", "1119562", "6235360", "2839336",
"6061457", "1683999", "14120418"],} \\
{\hspace*{2em}"method": ["12418427", "13938168", "201636367",
"53079390", "19172367", "1119562", "2839336",
"6061457", "54436148", "1683999"],} \\
{\hspace*{2em}"result": ["248887396", "25508712", "235418915",
"231786409", "233839149", "1119562", "2839336",
"6061457", "53554040", "1683999"]} \\
{\hspace*{1em}\},} \\

{\hspace*{1em}"questions": \{} \\
{\hspace*{2em}"background": "What are the limitations of current
prepositional phrase attachment resolution methods and how can
they be improved through representation learning and structured
learning approaches?",} \\
{\hspace*{2em}"method": "Identify innovative methods in PP
attachment, lexical feature representation, and unsupervised
word alignment models.",} \\
{\hspace*{2em}"result": "Examine parsers achieving superior
performance over SVM classifiers and traditional parsers in
diverse languages and tasks."} \\
{\hspace*{1em}\},} \\

{\hspace*{1em}"metadata": \{} \\
{\hspace*{2em}"2621465": \{} \\
{\hspace*{3em}"title": "Exploring Compositional Architectures
and Word Vector Representations for Prepositional Phrase
Attachment",} \\
{\hspace*{3em}"abstract\_sentences": [} \\

{\hspace*{4em}"Prepositional phrase (PP) attachment
disambiguation is a known challenge in syntactic parsing.",} \\

{\hspace*{4em}"The lexical sparsity associated with PP
attachments motivates research in word representations that
can capture pertinent syntactic and semantic features of
the word.",} \\

{\hspace*{4em}"One promising solution is to use word vectors
induced from large amounts of raw text.",} \\

{\hspace*{4em}"However, state-of-the-art systems that employ
such representations yield modest gains in PP attachment
accuracy.",} \\

{\hspace*{4em}"In this paper, we show that word vector
representations can yield significant PP attachment
performance gains.",} \\

{\hspace*{4em}"This is achieved via a non-linear architecture
that is discriminatively trained to maximize PP attachment
accuracy.",} \\

{\hspace*{4em}"The architecture is initialized with word vectors
trained from unlabeled data, and relearns those to maximize
attachment accuracy.",} \\

{\hspace*{4em}"We obtain additional performance gains with
alternative representations such as dependency-based
word vectors.",} \\

{\hspace*{4em}"When tested on both English and Arabic datasets,
our method outperforms both a strong SVM classifier and
state-of-the-art parsers.",} \\

{\hspace*{4em}"For instance, we achieve 82.6\% PP attachment
accuracy on Arabic, while the Turbo and Charniak self-trained
parsers obtain 76.7\% and 80.8\% respectively."} \\

{\hspace*{3em}],} \\
{\hspace*{3em}"facet\_labels":
["bg", "bg", "bg", "bg", "mt", "mt", "mt", "rs", "rs", "rs"]} \\
{\hspace*{2em}\},} \\

{\hspace*{2em}"368524": \{"title": "...",
"abstract\_sentences": ["..."], "facet\_labels": ["..."]\},} \\

{\hspace*{2em}"6946103": \{"title": "...",
"abstract\_sentences": ["..."], "facet\_labels": ["..."]\},} \\

{\hspace*{2em}...} \\
{\hspace*{1em}\}} \\
{\}} \\
\bottomrule
\end{tabular}
\end{table*}

\end{document}